\documentclass[preprint,authoryear,12pt]{elsarticle} 

\usepackage{color}
\usepackage{fullpage}
\usepackage{amsmath}
\usepackage{mathrsfs}
\usepackage[table]{xcolor}
\usepackage{newfloat}
\usepackage{graphicx}
\usepackage{subfig}
\usepackage{multirow}
\usepackage{caption}
\usepackage{arydshln}
\usepackage{tikz}
\usetikzlibrary{arrows,positioning,calc}
\usepackage{pstool}
\usepackage{textcomp}
\usepackage[hidelinks]{hyperref}
\hypersetup{colorlinks,bookmarksopen,linkcolor=blue,citecolor=blue,urlcolor=blue}

\newcommand{\bs}[1]{{\boldsymbol{#1}}}

\newcommand{\tr}[1]{\mathrm{tr\,{#1}}}

\newcommand{\ignore}[1]{}
\DeclareFloatingEnvironment{algorithm}

\usepackage{amssymb}

\journal{Mechanics of Materials}

\begin{document}

\begin{frontmatter}



\title{Size effects in nonlinear periodic materials exhibiting reversible pattern transformations\tnoteref{titlefoot}}


\author[affiliation]{M.M.~Ameen}
\ead{M.Ameen@tue.nl}

\author[affiliation]{O.~Roko\v{s}}
\ead{O.Rokos@tue.nl}

\author[affiliation]{R.H.J.~Peerlings\corref{mycorrespondingauthor}}
\cortext[mycorrespondingauthor]{Corresponding author.}
\ead{R.H.J.Peerlings@tue.nl}

\author[affiliation]{M.G.D.~Geers}
\ead{M.G.D.Geers@tue.nl}

\address[affiliation]{Mechanics of Materials, Department of Mechanical Engineering, Eindhoven University of Technology, P.O.~Box~513, 5600~MB~Eindhoven, The~Netherlands}

\tnotetext[titlefoot]{The post-print version of this article is published in \emph{Mech. Mater.}, \href{https://www.sciencedirect.com/science/article/pii/S0167663617307159}{10.1016/j.mechmat.2018.05.011}.}

\begin{abstract}
This paper focuses on size effects in periodic mechanical metamaterials driven by reversible pattern transformations due to local elastic buckling instabilities in their microstructure. Two distinct loading cases are studied: compression and bending, in which the material exhibits pattern transformation in the whole structure or only partially. The ratio between the height of the specimen and the size of a unit cell is defined as the scale ratio. A family of shifted microstructures, corresponding to all possible arrangements of the microstructure relative to the external boundary, is considered in order to determine the ensemble averaged solution computed for each scale ratio. In the compression case, the top and the bottom edges of the specimens are fully constrained, which introduces boundary layers with restricted pattern transformation. In the bending case, the top and bottom edges are free boundaries resulting in compliant boundary layers, whereas additional size effects emerge from imposed strain gradient. For comparison, the classical homogenization solution is computed and shown to match well with the ensemble averaged numerical solution only for very large scale ratios. For smaller scale ratios, where a size effect dominates, the classical homogenization no longer applies.
\end{abstract}


\end{frontmatter}


%
%
\section{Introduction}
\label{Sect.Introduction}
\begin{figure}
	\centering
	\subfloat[reference]{\includegraphics[scale=0.5]{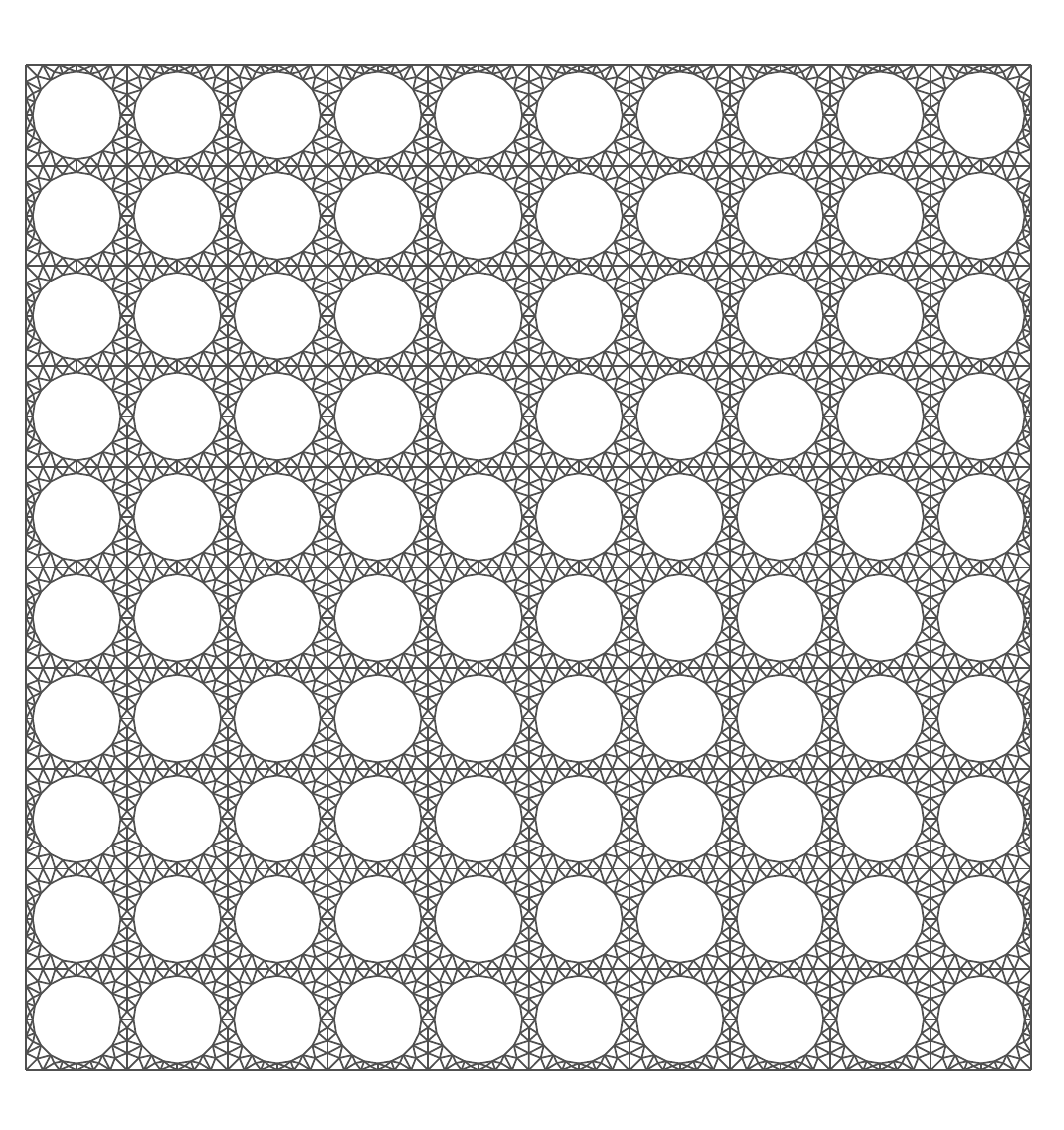}\label{Figure.Sketch1a}}\hspace{1.0em}
	\subfloat[deformed]{\includegraphics[scale=0.5]{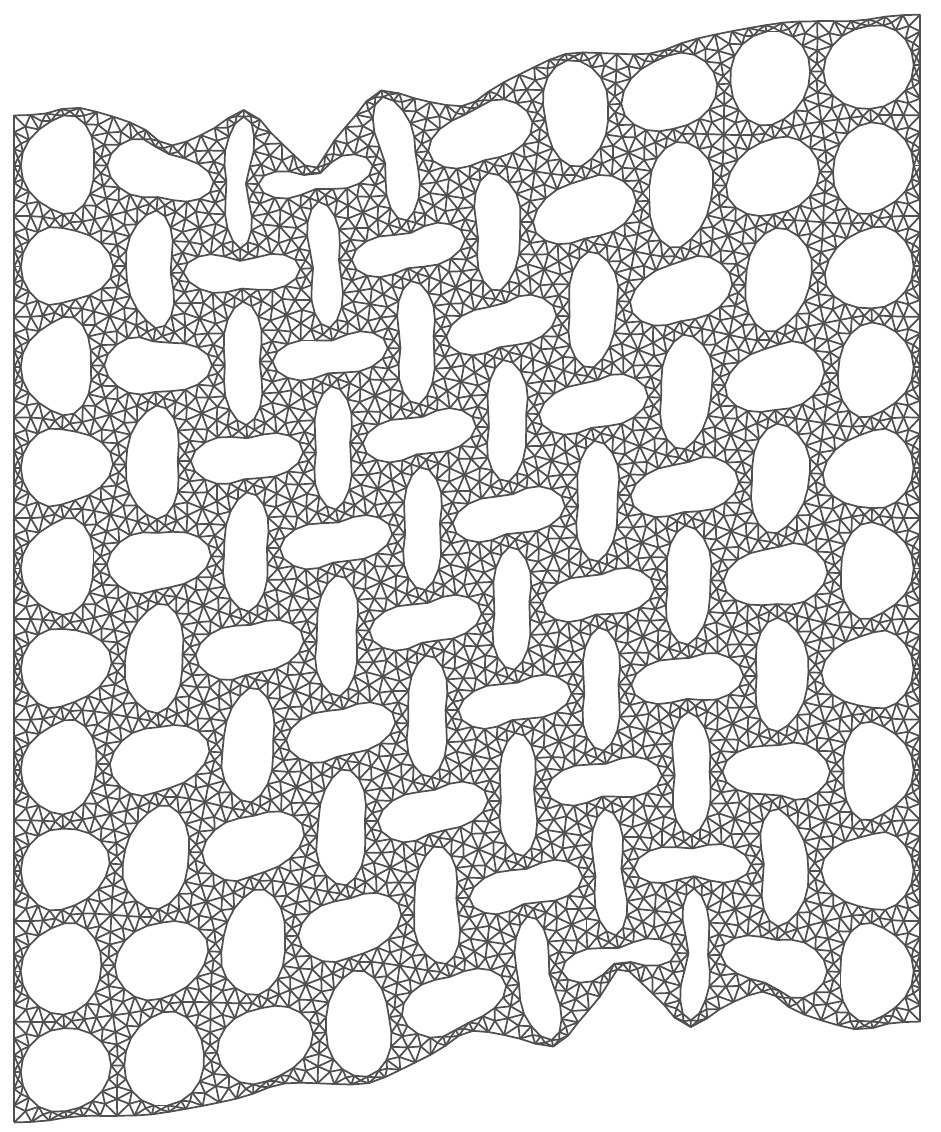}\label{Figure.Sketch1b}}
  	\caption{An elastomeric material containing periodically arranged holes is subjected to a shear deformation along the two lateral edges and a uniform compression along the horizontal direction. The undeformed geometry is shown on the left and the deformed geometry on the right.}
  \label{Figure.Sketch1}
\end{figure}

Over the past decade, cellular materials have found widespread use in thermal, mechanical, and acoustic applications. Their mechanical response is largely influenced by the geometry at the scale of the microstructure. This has inspired researchers to design microstructures which exhibit anomalous behavior \citep[see][]{florijn2014}. In particular, architectured cellular materials consisting of periodically arranged circular holes in an elastomer base material exhibit a pattern transformation, which is triggered when the applied compressive load reaches a critical value. As a result of this transformation, the incremental effective properties of the material change dramatically. Fig.~\ref{Figure.Sketch1} shows the deformation pattern of such a material under combined compression and shear.

One of the earlier works on such porous elastomers is presented by~\cite{Mullin2007a}, where it is shown that pattern transformations are triggered by a reversible elastic instability. The onset of instabilities in materials with arbitrary microstructures for finitely strained, rate-independent solids can be modelled through Bloch analysis \citep[see][]{geymonat1993}. This approach was later extended also to periodic, microporous, elastoplastic coatings in the work of \cite{Singamaneni2009a}, where the authors showed that periodic cross-linked microstructures provide the ability to lock in the transformed pattern with complete relaxation of the internal stresses. 

Because transforming porous elastomers exhibit very different post-buckling behavior, they are categorized as mechanical metamaterials. Auxetic behavior of such materials is explored in more detail by~\cite{Bertoldi2010a}. Materials with empty holes or holes filled with hydrogel have been studied by~\cite{Hu2013}, where the influence of the shapes and geometry of the holes on the mechanical properties was demonstrated. Besides the mechanical properties, their optical and acoustic properties in the post-bifurcation regime are active areas of research. 
\cite{Bertoldi2008d} studied the phononic behavior of periodic elastomeric solids and showed that the pattern transformations can give rise to reversible and repeatable phononic band gaps. \cite{Krishnan2009} studied the optical properties of 2-D elastomeric photonic crystals to design strain-tunable optomechanical materials for sensing and actuation. \cite{Singamaneni2010a} reviewed the applications of buckling instabilities in elastic solids, indicating their potential application in stretchable electronics, phononic crystals, and tunable surfaces, whereas~\cite{Liu2016a} explored negative swelling behavior in soft architected materials. 

Materials whose response to uniaxial compression can be programmed by lateral confinement are called \textit{programmable mechanical materials}, a notion introduced by~\cite{florijn2014}. \cite{Shan2014a} explored the behavior of such elastomeric materials consisting specifically of staggered arrays of holes. The authors showed that multiple folding mechanisms arise, which can be used to tune the dynamic properties in phononic crystals. \cite{Yang2015} further utilized the nonlinear reversible collapse behavior of an elastomeric structure consisting of elastic beams and interconnected cavities to study various motions, which can be exploited in soft robotics. \cite{Coulais2016a} studied the case of elliptic holes and showed that the material nonlinearity becomes large in the limit of large and near-circular holes. 

The mechanical behaviour exhibited by pattern transforming elastomers is significantly influenced by applied boundary conditions. Recall Fig.~\ref{Figure.Sketch1}, where an example of a square specimen consisting of~$ 10 \times 10 $ holes, subjected to uniform compression along the horizontal direction and a shear along the vertical direction, is shown. The top and bottom boundaries are left free, whereas both lateral edges are constrained in horizontal as well as vertical direction. These constrained edges introduce boundary layers while a band of pattern-transformed region localizes in one of the diagonal directions due to applied load. These effects strongly depend on the size of the holes relative to the size of the specimen, which is a typical \textit{size effect}. In the available literature, various kinds of size effects have been studied extensively. For instance, \cite{arzt1998} gives a comparative review of size effects in materials due to dimensional constraints. \cite{brezny1990} studied a reticulated vitreous carbon foam and showed experimentally that the compressive and bending strength of such materials is inversely proportional to the cell size. \cite{frantziskonis1997} studied boundary effects in 1-D bars subjected to uniaxial loads. \cite{onck2001size}
studied size effects in metallic foams subjected to various mechanical loading triggered by plastic buckling in the cell. They showed that for compression, the effective macroscopic strength and stiffness of such foams decreases with decreasing sample size. Such a behavior was attributed to the relative increase in thickness of the weak boundary layers located at the stress-free edges.
\cite{chen2002size} studied aluminum foams sandwiched between metallic substrates. They showed that under constrained deformation, the yield strength increases by almost a factor of two as the height to width ratio of the foam is decreased from~$20$ to~$3$. It was also reported that regular hexagonal honeycomb foams do not show this type of size effect. 
  
In contrast to the above listed works focusing mainly on foams and honeycombs, neither qualitative nor quantitative study of size effects in transforming porous elastomers (and metamaterials based hereon) has been provided in the literature. In order to fill this gap, we provide in this paper a systematic study of precisely those kinds of microstructures and their size effects. In particular, a regular periodic mechanical metamaterial which undergoes reversible pattern transformations is considered. Transformed patterns induce a characteristic fluctuation in the deformation field, which is responsible for its anomalous behavior. A hyperelastic material containing periodically arranged circular holes is adopted. Two distinct loading cases are studied under plane strain conditions: compression and bending, in which the material exhibits pattern transformation in the whole structure or part of the structure respectively. The effect of boundary conditions on the size effect is explored in detail through fully constrained top and bottom boundaries (in the case of compression) and free boundaries (in the case of bending). The ratio between the height of the specimen~$L$ and the size of a unit cell~$\ell$ is defined as the scale ratio~$L/\ell$. Since the exact position of the microstructure relative to the specimen geometry cannot be fully controlled in most practical applications, all possible translations of the microstructure are considered for each scale ratio. The ensemble average of this entire family of solutions is then used to define statistically representative macroscopic response. For the compression case, the constrained boundaries introduce boundary layers in which the pattern transformation is restricted. As~$L/\ell$ approaches unity, the relative thickness of these boundary layers increases, giving rise to a strong size effect. For the bending case, the applied load induces a strain gradient resulting in a partial transformation of the pattern, where the upper (tensile) half of the specimen remains untransformed. For comparison, the conventional RVE-based computational homogenization solution (FE\textsuperscript{2}) is reported. It is shown that good agreement with the ensemble averaged full-scale numerical simulation is achieved only for rather large scale ratios. For smaller values of~$L/\ell$, significant deviations occur, indicating strong size dependence. Extracting the ensemble averaged solution as a function of the scale ratio allows to identify the scale separation limit beyond which conventional homogenization starts to deviate from the exact solution. A quantitative study on the size effect is also performed by analyzing the role of the boundary layer thickness. 

The contents of this paper is divided into three sections. The first one defines the problem to be studied, including its geometry, material properties, and the boundary conditions used for the two loading cases. It also details the ensemble averaging scheme used to define the homogenized solution for a range of scale ratios, and the numerical implementation of the model by finite element method. The next section, Section~\ref{Section.Results}, reports the detailed results obtained for the case of compression and bending. Finally, our paper closes with a summary and conclusions in Section~\ref{Section.SummaryConclusion}. For completeness, \ref{Section:A} summarizes basic ideas of the first-order computational homogenization. 

Throughout this paper, the following notation conventions are used:
\begin{itemize}
\itemsep0em 

\item[--] scalars: $ a $

\item[--] vectors: $ \vec{a} $

\item[--] second-order tensors: $ \bs{A} $

\item[--] fourth-order tensors: $ \bs{A}^4 $

\item[--] matrices: $ \bs{\mathsf{A}} $

\item[--] $ \vec{a} \cdot \vec{b} = a_i b_j $

\item[--] $ \bs{A} \cdot \vec{b} =  A_{ij} b_j\vec{e}_i $

\item[--] conjugate: $ \bs{A}^c, A_{ij}^c = A_{ji} $

\item[--] gradient operator: $ \displaystyle \nabla \vec{a} = \frac{\partial a_j}{\partial x_i} \vec{e}_i \vec{e}_j $

\item[--] divergence operator: $ \displaystyle \nabla \cdot \vec{a} = \frac{\partial a_i}{\partial x_i} $.

\end{itemize}
%
%
\section{Problem description and methodology}
\label{Section.ProblemDescription}
In the following subsections, employed geometry, material properties, and boundary conditions are detailed. The material and geometry used in both examples are inspired by the experimental \textit{Specimen~1} reported in the work of \cite{Bertoldi2008d}. 
%
%
\subsection{Geometry}
\label{SubSect.Geometry}
\begin{figure}[h]
	\centering
	\subfloat[compression]{\includegraphics[scale=1]{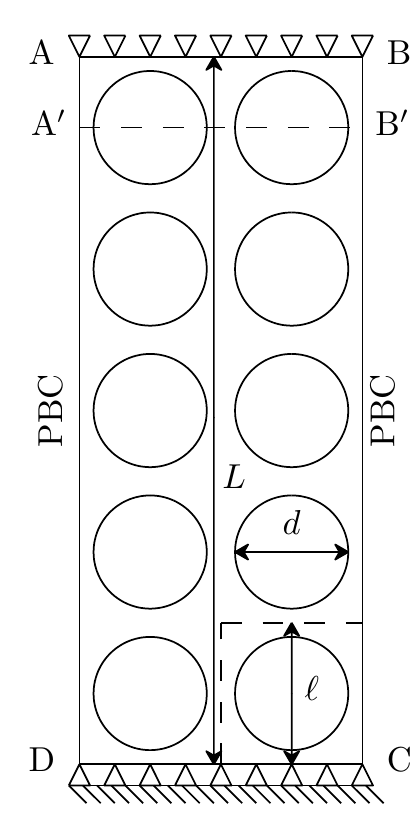}\label{Figure.Sketch2a}}\hspace{1.0em}
	\subfloat[bending]{\includegraphics[scale=1]{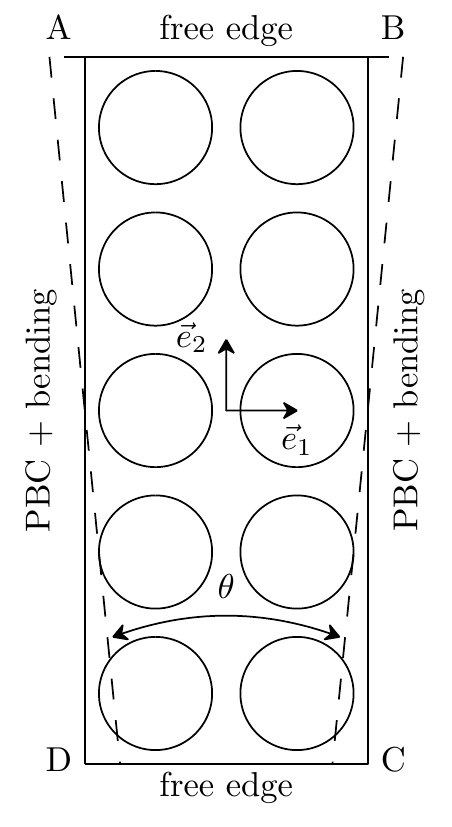}\label{Figure.Sketch2b}}	
 	\caption{Sketch of the specimens subjected to the two loading conditions for~(a) uniaxial compression and~(b) bending. Note that the problem domain is defined as~$\Omega^\mathrm{P} = (-\infty,\infty)\times(-L/2,L/2)$, whereas the model domain is limited only to~$\Omega = (-\ell,\ell)\times(-L/2,L/2)$ due to periodicity.}
	\label{Figure.Sketch2}
\end{figure}

The problem considered consists of a two-dimensional, infinitely wide, periodic, and heterogenous hyperelastic medium subjected to two distinct loading conditions, see Fig.~\ref{Figure.Sketch2}. The size of a single unit cell, $\ell$, gives the microstructural length scale, whereas the specimen height, $L$, is identified as the macroscopic length scale. In particular, according to~\cite{Bertoldi2008d}, we choose unit cell size~$\ell = 9.97$~mm, diameter of holes~$d = 8.67$~mm, and will consider for further purposes only integer number of unit cells per specimens' height, i.e.~$L/\ell \in \mathbb{N}$. In total~$ 25 $ distinct scale ratios will be used, varied such that~$ 4 \leq L/\ell \leq 128 $. In all cases, the resulting deformation limits to a~$ 2 \times 2$ repetitive pattern (cf. Fig.~\ref{Figure.Sketch1}), allowing us to model only a periodic,~$2 \ell \times L$ large, part of otherwise infinitely wide domain. Throughout the paper, a Cartesian coordinate system is used with its origin at the midpoint of the reduced \textit{model domain}~$\Omega = (-\ell,\ell)\times(-L/2,L/2)$ with boundary~$\partial\Omega$, whereas the actual \textit{problem domain} is~$\Omega^\mathrm{P} = (-\infty,\infty)\times(-L/2,L/2)$ with boundary~$\partial\Omega^\mathrm{P}$.
%
%
\subsection{Material behavior}
\label{SubSect.MatBehaviour}
%
The elastomer base material is a hyperelastic material described by a two-term Mooney-Rivlin model
whose strain energy density function~$W$ is given as
\begin{equation}
W( \bs{F} (\vec{X}) ) = m_1 (I_1 - 3) + m_2 (I_1 - 3)^2 - 2 m_1 \log{J} + \frac{1}{2} K (J-1)^2.
\label{Eq:1}  
\end{equation}
In this equation, $\vec{X} \in \Omega$, $\vec{X} = X_1\vec{e}_1 + X_2\vec{e}_2$, is the position vector in the plane of the problem for the undeformed geometry. $ \bs{F} = ( \nabla_0 \vec{x} )^c $ defines the deformation gradient where~$ \vec{x} $ gives the current position vector and~$\nabla_0$ denotes the gradient operator with respect to the reference coordinate frame. $ J $ denotes the determinant of~$ \bs{F} $, and~$I_1$ and~$I_2$ are the invariants of the right Cauchy-Green deformation tensor~$ \bs{C} = \bs{F}^c \cdot \bs{F} $, given as
\begin{align}
I_1 &= \tr{\bs{C}} \label{Eq:2}, \\
I_2 &= \frac{1}{2} \left[(\tr{\bs{C}})^2 - \tr{\bs{C}^2} \right] \label{Eq:3}.
\end{align}

The adopted material parameters are given by~$ m_1 = 0.55~$MPa, $ m_2 = 0.3~$MPa, and bulk modulus~$K = 55$~MPa, following again the experimental characterization by~\cite{Bertoldi2008d}.
%
%
\subsection{Boundary conditions}
\label{BC}
In Fig.~\ref{Figure.Sketch2}, the applied boundary conditions are schematically shown. For the compression case, the lateral edges~AD and~BC are subjected to standard periodic boundary conditions to model the infinitely wide geometry. These conditions can be expressed as
\begin{equation}
\vec{x}_\mathrm{BC} - \vec{x}_\mathrm{AD} = \vec{X}_\mathrm{BC} - \vec{X}_\mathrm{AD},
\label{standard_PBC}
\end{equation}
where~$\vec{X}_\mathrm{BC}$ and~$\vec{x}_\mathrm{BC}$ denote position vectors of corresponding material points located on the edge BC in the reference and deformed configuration (similar definition holds for the edge~AD); for further details see e.g.~\cite{Kouznetsova:2001}. The bottom edge CD is fully constrained, i.e. fixed in both horizontal and vertical directions. The top edge AB is fixed in the horizontal direction and is subjected to a uniform (negative) displacement in the vertical direction. These fixed boundaries first introduce compressive strain throughout the specimen height~$L$, and later also boundary layers in which pattern transformation is restricted. The emergence of boundary layers and the influence of their thicknesses with varying scale ratio is of critical interest in this work. The vertical reaction force at the bottom edge CD is measured for the applied compressive strain.

For the bending case, the lateral edges~AD and~BC are subjected to a relative rotation~$\theta$, combined with periodic boundary conditions, as indicated schematically in Fig.~2b. Mathematically, this can be written as
\begin{equation}
\vec{x}_\mathrm{AD} = \bs{R}(\theta) \cdot \vec{x}_\mathrm{BC},
\label{rotated_PBC}
\end{equation}
where~$\bs{R}(\theta)$ is the rotation tensor. The material points in the deformed configuration~$\vec{x}_\mathrm{AD}$ and~$\vec{x}_\mathrm{BC}$ are expressed relative to the center of rotation~$\vec{P}$, e.g.~$\vec{P} = (0,-P_{X_2})^\mathsf{T}$. For implementation purposes, however, and in order to eliminate the dependence on~$\vec{P}$, Eq.~\eqref{rotated_PBC} is, after discretization, written for each node located on the edges~BC and~AD, and subsequently expressed relatively with respect to each other by pairwise subtraction. The top and bottom edges, AB and CD, are traction free. To eliminate rigid body displacements and for further analyses (see Section~\ref{Bending}), two points located at~$(\pm\ell,0)$ are fixed in both directions. These constraints effectively fix the vertical positioning of the neutral axis, resulting in bending combined with normal force. The applied angle introduces partial patterning, where the upper half of the specimen remains untransformed. Near the two free boundaries, the effect of the compliant boundary layers can be observed. For characterizing the mechanical response, the resulting bending moment will be quantified.
%
%
\subsection{Ensemble averaging}
\label{Homogenization}
Because the exact position of the microstructure with respect to the macroscopic specimen cannot be fully controlled in most practical applications, all possible translations of the microstructure are considered and averaged to recover a statistically representative response. Whereas the microstructural morphology is fixed in space, its position can randomly vary in space according to a uniform probability distribution.
This is based on the definition by~\cite{Smyshlyaev}, and demonstrated in our earlier work (see \cite{Ameen2018}).

\begin{figure}
	\centering
	\subfloat[$\zeta_v = 0$]{\includegraphics[scale=0.5]{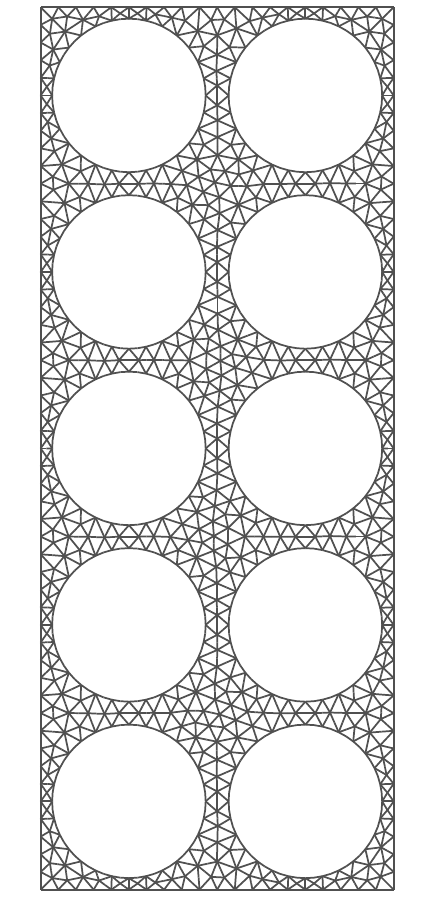}\label{Figure.shiftsa}}\hspace{0.5em}
	\subfloat[$\zeta_v = 0.25$]{\includegraphics[scale=0.5]{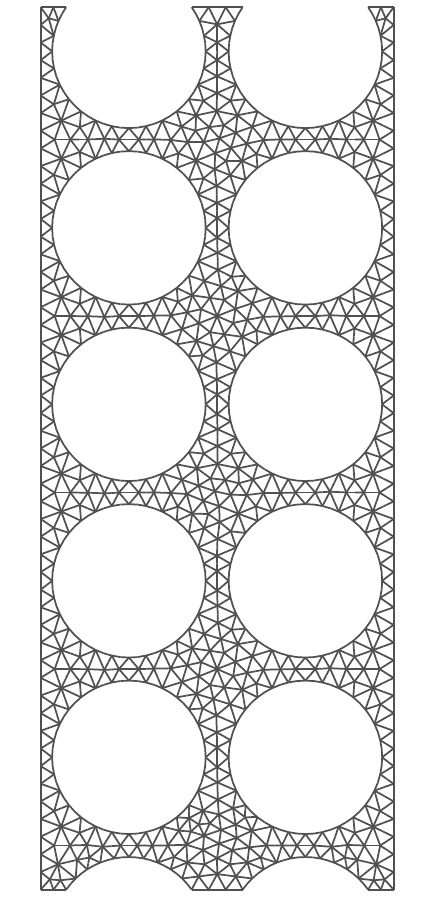}\label{Figure.shiftsb}}\hspace{0.5em}
	\subfloat[$\zeta_v = 0.5$]{\includegraphics[scale=0.5]{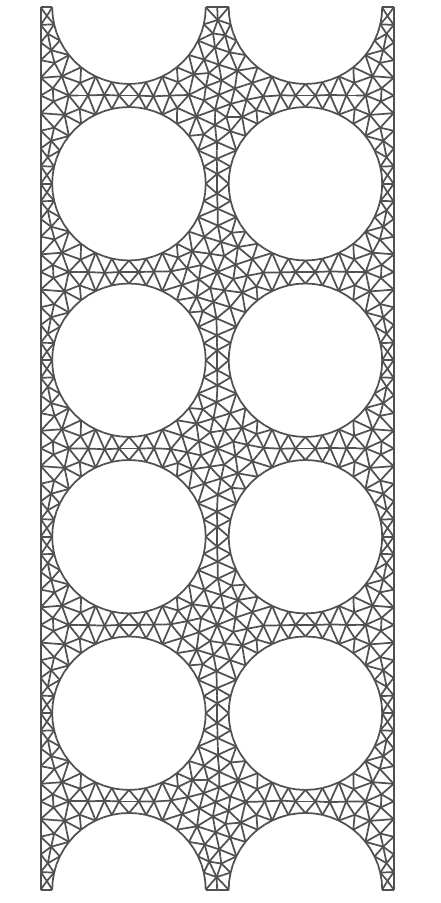}\label{Figure.shiftsc}}
	\caption{Specimens with scale ratio~$ L/\ell = 5 $ and different positions of the microstructure relative to the top and bottom boundaries.}
	\label{Figure.shifts}
\end{figure}

To describe the relative positioning of the microstructure with respect to the specimen geometry, we introduce a constant translation vector~$\vec{\zeta} \in \Omega$,
\begin{equation}
\vec{\zeta} = \zeta_h \, \ell \, \vec{e}_1 + \zeta_v \, \ell \, \vec{e}_2, \quad -1 \leq \zeta_h < 1, \quad -L/(2\ell) \leq \zeta_v < L/(2\ell),
\label{Eq:shift}
\end{equation}
where~$\zeta_h$ and~$\zeta_v$ denote normalized shifts along the horizontal and vertical directions relative to the unit cell size~$\ell$. Since the two loading cases considered are periodic in the horizontal direction, horizontally shifting the microstructure has no effect other than shifting the solution correspondingly. As a consequence, statistical ensemble average over all horizontal shifts can effectively be obtained as a simple spatial average, i.e.
\begin{equation}
\mbox{mean } P_{11}^{\zeta_v} (X_2) = \frac{1}{2\ell} \int_{-\ell}^{\ell}P_{11}^{\zeta_v}(X_1,X_2)\,\mathrm{d}X_1, \quad -L/2 \leq X_2 \leq L/2,
\label{Eq:meanP}
\end{equation}
which is independent of the horizontal coordinate~$X_1$. In Eq.~\eqref{Eq:meanP}, the~$P_{11}$ component of the first Piola--Kirchhoff stress tensor~$\bs{P}$ has been used as an example. Regarding the vertical shifts~$\zeta_v$, it is sufficient to consider only~$ 0 \leq \zeta_v < 1 $ due to the periodicity of the microstructure in the vertical direction with the period~$\ell$. The vertical shifts need to be, however, accounted for computationally, and an example of three realizations out of the entire family of vertical shifts considered for the scale ratio~$ L/\ell = 5 $ is shown in Fig.~\ref{Figure.shifts}. The final ensemble averaged quantities, averaged over shifts along both directions and associated with each scale ratio~$L/\ell$, are obtained as
\begin{equation}
\mbox{ensemble average } P_{11} (X_2) = \int_0^1 \mbox{mean }P_{11}^{\zeta_v}(X_2)\,\mathrm{d}\zeta_v, \quad -L/2 \leq X_2 \leq L/2,
\label{Eq:ensembleAv}
\end{equation}
where the~$P_{11}$ stress component has been used as an example again. 

All quantities reported below such as the first Piola--Kirchhoff stress~$\bs{P}$ or the deformation gradient~$\bs{F}$ will be for better clarity provided for individual vertical shifts~$\zeta_v$ in terms of their horizontal averages defined in Eq.~\eqref{Eq:meanP}. To this end, a fixed regular grid of (sampling) points is introduced in the problem domain~$\Omega$ with a spacing~$ \ell/100 $ along both~$ \vec{e}_1 $ and~$ \vec{e}_2 $ directions, and the corresponding integrals are always carried out numerically. Inside the holes, a linear elastic material behavior is assumed, solely for the purpose of computing the displacements and strains in these regions. Here, the stresses equal zero, the displacements are interpolated using the linear elastic material response, and strains are computed as displacement gradients. Note that the interpolation is performed as a post-processing step, and all DNS simulations are carried out with the holes. This interpolation step allows to construct all quantities in the entire model domain~$ \Omega $, from which their averages can be derived. The shift interval~$\zeta_v \in [0, 1)$ is uniformly discretized into~$n_{\zeta_v}(L/\ell)$ points. For each scale ratio, the number of shifts considered is 
\begin{equation}
n_{\zeta_v}(L/\ell) = \mbox{nint}\left[200 - \frac{180}{128-4}(L/\ell - 4)\right],
\label{Eq:numshifts}
\end{equation}
where~$\mbox{nint}[\bullet]$ denotes the nearest integer to~$\bullet$. Note that ensemble averaging does not introduce any new length scales, unlike methods like moving volume averaging, which makes it an apt method especially for the cases with relatively small~$ L/\ell $ ratios.
%
%
\subsection{Numerical Implementation}
\label{model}
The numerical simulations are carried out using a nonlinear finite element code which makes use of the Total Lagrangian formulation, see e.g.~\cite{crisfield2012}. The finite element mesh is constructed out of six-noded, quadratic triangular elements with three Gauss integration points. The finite element mesh generator \textit{gmsh} \citep[see][]{NME2579} is used for the discretization. A rigorous mesh refinement study has been performed to establish an accurate discretization, which resulted in a typical element size~$h_\mathrm{FE} = \ell/10$. The model was furthermore validated by the experimental results reported by~\cite{Bertoldi2008d}. All full-scale numerical solutions are computed for the entire family of translated microstructures considered, for the two loading cases, and for each scale ratio~$ L/\ell $. In all numerical simulations, a total of~$200$ load increments are used.

Under compression, the microstructure exhibits local instabilities initiated by internal buckling. These instabilities trigger the pattern transformation, upon which the material behavior changes from an approximately linear initial response to a complex post-buckling response, depending on the geometry and type of boundary conditions applied. The instability is detected using pivot checking, see e.g.~\cite{Wriggers2008b}. When a negative pivot is encountered, the Newton solver is initialized towards the lowest mode by adding a linear perturbation
\begin{equation}
\Delta\vec{u} = w \,\ell \, \frac{\vec{\phi}_{1}}{\|\vec{\phi}_{1}\|_\infty}
\label{Eq:perturbation}
\end{equation}
to the current solution~$\vec{u}$. In Eq.~\eqref{Eq:perturbation}, $\vec{\phi}_1$ denotes the eigenmode corresponding to the lowest eigenvalue of the current tangential stiffness matrix, and~$w \in [1 \cdot 10^{-4}, 1 \cdot 10^{-2}]$ is the scaling factor that is changed adaptively. The algorithm is further equipped with automatic restart and damping. Path-following methods such as the arc-length method are not strictly required as no structural softening nor snap-back occur, cf. Section~\ref{Section.Results} below. The above-described strategy allows for simple yet efficient tracking of the mechanical response without introducing any artificial imperfections, that could otherwise pollute the observed size effects.
%
%
\section{Results}
\label{Section.Results}
In this section, computed results are presented separately for the two loading cases, uniaxial compression and bending. First, individual shifted solutions are introduced and discussed, followed by ensemble averaged solutions and their behavior across the scales~$L/\ell$.
%
%
\subsection{Uniaxial compression}
\label{UniaxCompression}
The maximum nominal applied linear strain attains~$ u/L = 0.1 $ (where~$ u $ is the applied vertical displacement), a value that is based on the experimental results from the work of \cite{Bertoldi2008d} and numerical simulations for various scale ratios. 
This value ensures that specimens of all scale ratios of interest, $4\leq L/\ell \leq 128$, are subjected to pattern transformations without deforming in a higher mode (destroying thus the patterns).

\begin{figure}
	\centering
	\subfloat[deformed configurations]{\includegraphics[scale=1]{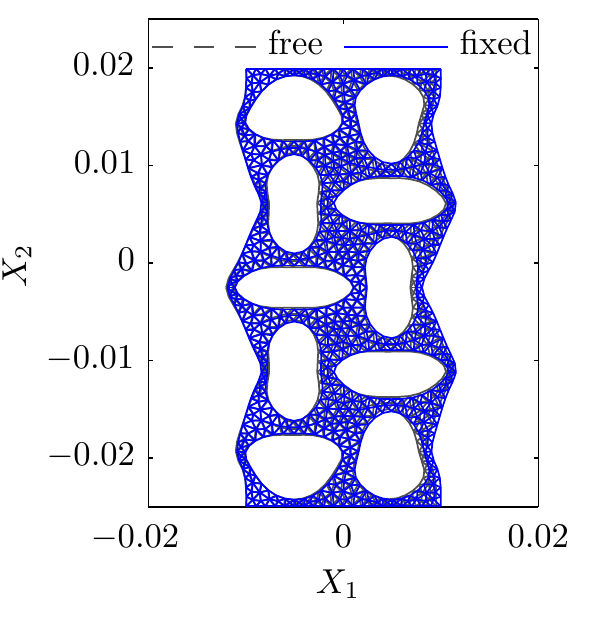}\label{Figure.fixedvsfreea}}
	\subfloat[stress--strain diagram]{\includegraphics[scale=1]{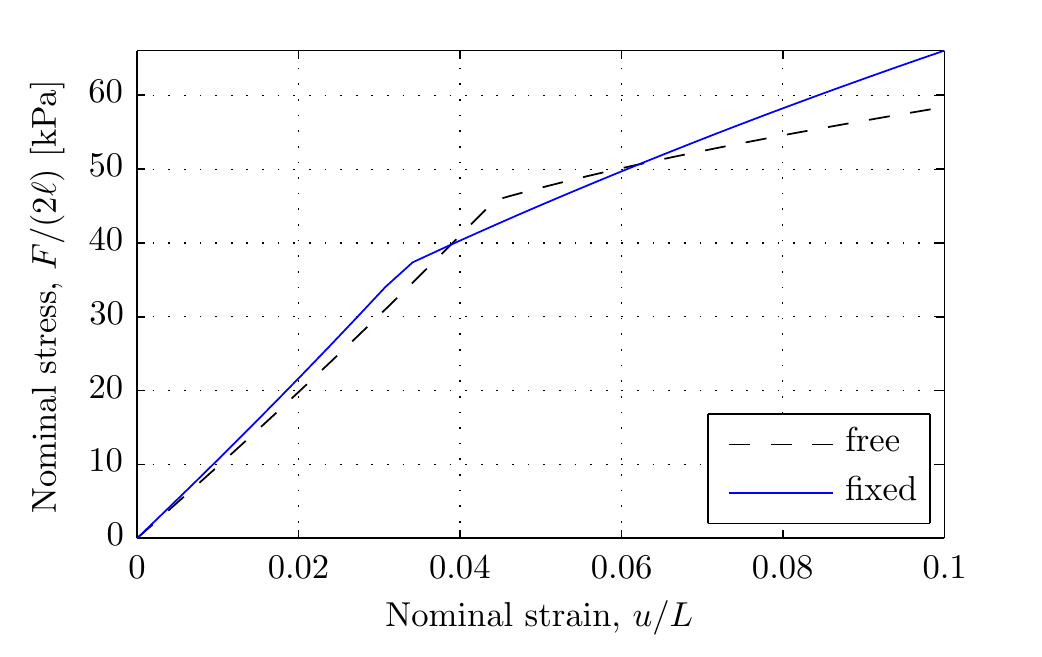}\label{Figure.fixedvsfreeb}}	 
	\caption{Comparison between free vs fixed horizontal displacement at the top and bottom edges for a specimen with scale ratio~$L/\ell = 5$ and~$\zeta_v = 0$, subjected to uniform compression. The deformed shapes are shown on the left, whereas the nominal stress vs the applied nominal strain is on the right.}
	\label{Figure.fixedvsfree}
\end{figure}

A comparative study is first performed in order to assess two different boundary conditions, one in which the horizontal displacement at the top and bottom boundaries is fixed and another case where the two boundaries are allowed to freely expand/contract in the horizontal direction. The deformed shapes and corresponding stress--strain curves for the two cases are shown in Fig.~\ref{Figure.fixedvsfree} for the scale ratio~$L/\ell = 5$ and zero vertical shifts. In Fig.~\ref{Figure.fixedvsfreeb}, the nominal stress is plotted against the applied nominal strain. Here, the nominal stress~$ F/(2\ell) $ is defined as the ratio between the computed reaction force~$ F $ per unit thickness and the width of the model domain~$ 2\ell $, and the nominal strain~$ u/L $ is defined as the ratio between the vertical displacement~$ u $ and the specimen height~$ L $. The initial overall stiffness is higher for the fixed case, but it undergoes bifurcation at a lower nominal strain than the case with unconstrained horizontal deformation. The post-bifurcation stiffness is also higher for the former. The deformed shapes for the two cases, however, are quite similar. In the case of free horizontal displacements, local buckling of vertical ligaments in cases in which holes cut through horizontal boundaries may occur, which can trigger convergence issues. To prevent this, all further simulations for the case of uniaxial compression are performed with horizontally constrained top and bottom boundaries, i.e. the bottom edge is fully restrained in both directions and the top edge is restrained in the horizontal direction. Note that these boundary conditions differ from those used in experiments by~\cite{Bertoldi2008d}, where the finitely wide specimens were allowed to deform freely in the horizontal direction at the top and bottom edges. 
%
%
\subsubsection{Full-scale numerical solutions}
\label{DNS}
\begin{figure}
	\centering
	\mbox{}\hspace{2.0em}\subfloat[$\zeta_v = 0$]{\includegraphics[scale=0.5]{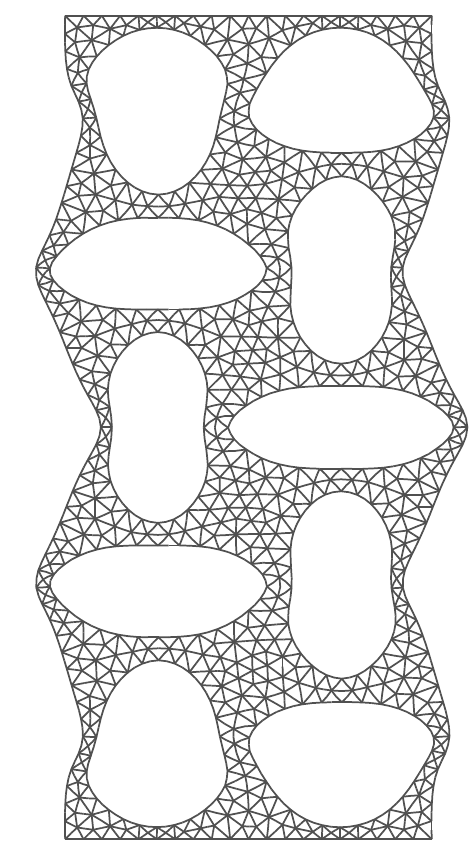}\label{Figure.shifts_compa}}\hspace{1.0em}
	\subfloat[$\zeta_v = 0.25$]{\includegraphics[scale=0.5]{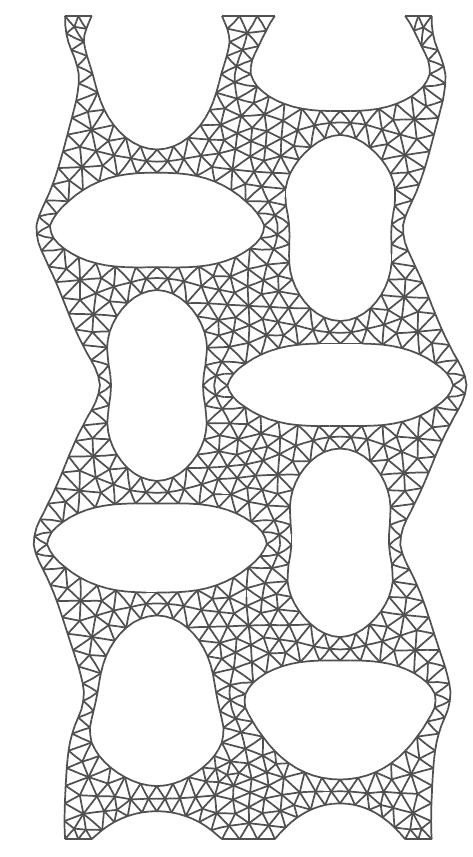}\label{Figure.shifts_compb}}\hspace{1.0em}
	\subfloat[$\zeta_v = 0.5$]{\includegraphics[scale=0.5]{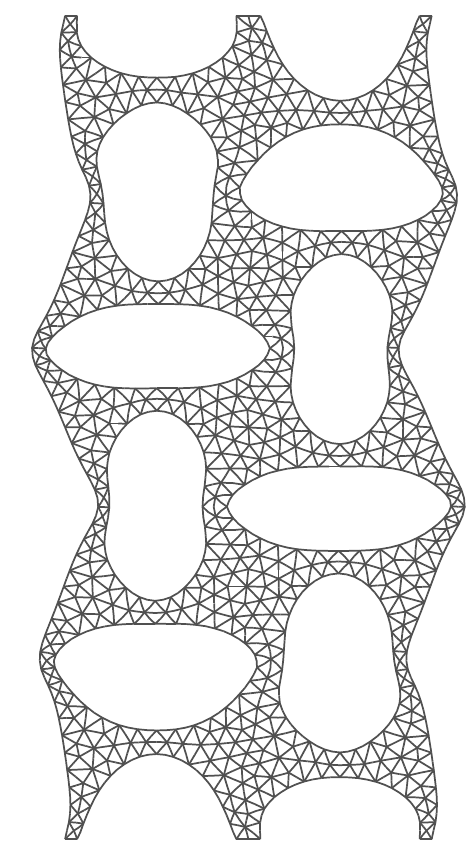}\label{Figure.shifts_compc}}\\
	\subfloat[stress--strain diagram]{\includegraphics[scale=1]{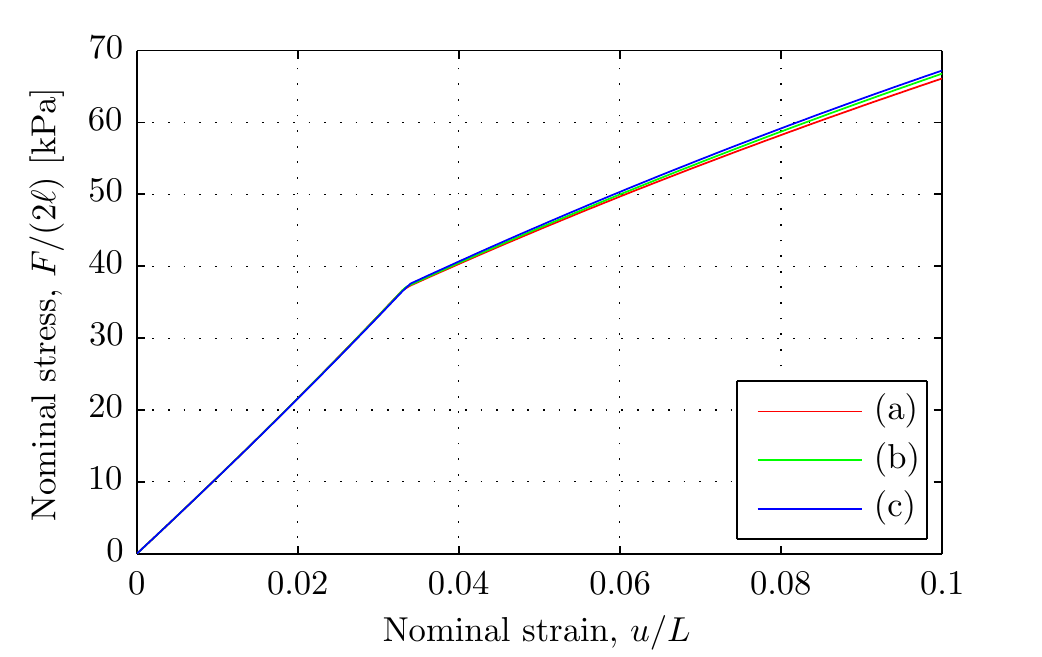}\label{Figure.shifts_compd}}	
	\caption{Deformed shapes under compression for scale ratio~$L/\ell = 5$ and different positions of the microstructure relative to the top and bottom boundaries: (a)~$\zeta_v = 0$, (b)~$\zeta_v = 0.25$ and~(c) $\zeta_v = 0.5$. The stress--strain curves for the three cases are shown in~(d).}
	\label{Figure.shifts_comp}
\end{figure}

Fig.~\ref{Figure.shifts_comp} shows the deformed shapes and the stress--strain curves for three different realizations (corresponding to the specimens shown in Fig.~\ref{Figure.shifts}) for scale ratio~$L/\ell = 5$. The influence of the vertical shift~$\zeta_v$ of the microstructure is clearly visible in the deformed shapes. Nevertheless, the stress--strain responses do not differ substantially, as can be observed in Fig.~\ref{Figure.shifts_compd}. 

Figs~\ref{Figure.shifts_compa} --~\ref{Figure.shifts_compc} clearly show that the entire specimen has undergone a pattern transformation, typical of such geometry. The circular holes are transformed into ellipses, with their major axis alternating along the~$ \vec{e_1} $ and~$ \vec{e_2} $ directions. This pattern provides a state of deformation with a lower free energy. Away from the top and bottom boundaries, the overall (average) deformation is more or less uniform, whereas close to the boundaries the circular holes do not transform into full ellipses, giving rise to boundary layers. 
%
%
\subsubsection{Influence of the scale ratio~$L/\ell$}
\label{ScaleRatio}
\begin{figure}
	\centering
	\subfloat[$L/\ell = 20$]{\includegraphics[scale=0.7,angle =90]{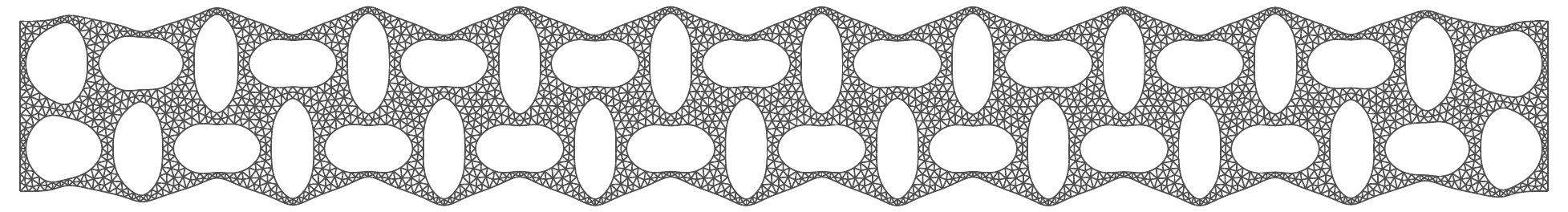}\label{Figure.sr_compa}}\\
	\subfloat[$L/\ell = 10$]{\includegraphics[scale=0.7,angle =90]{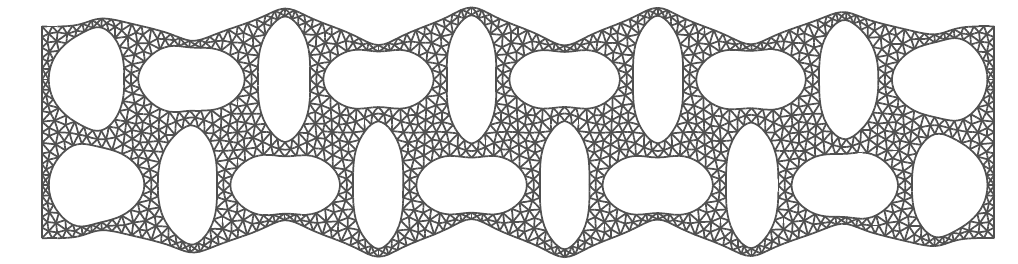}\label{Figure.sr_compb}}\hspace{2.0em}
	\subfloat[$L/\ell = 5$]{\includegraphics[scale=0.7,angle =90]{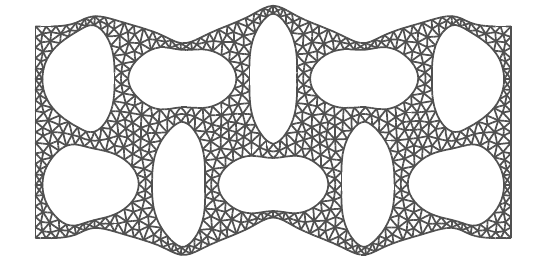}\label{Figure.sr_compc}}	
	\caption{Deformed shapes under compression (rotated~$90$\textdegree~anticlockwise) for specimens with zero vertical shift and three scale ratios.}
	\label{Figure.sr_comp}
\end{figure}

Fig.~\ref{Figure.sr_comp} shows the deformed shapes for three different scale ratios for the configuration~$\zeta_v = 0$. This shows the effect of boundary layers penetrating into the bulk of the specimens. For a relatively large scale ratio, $L/\ell = 20$ shown in Fig.~\ref{Figure.sr_compa}, except for a few layers of holes close to the boundaries all holes have transformed into ellipses. This in turn means that the bulk of the material undergoes a uniform overall macroscopic deformation for large scale ratios. For smaller scale ratios, $L/\ell = 10$ and~$L/\ell = 5$ shown in Figs~\ref{Figure.sr_compb} and~\ref{Figure.sr_compc}, the boundary layers dominate and only a smaller (or zero) region in the bulk undergoes uniform (macroscopic) deformation. 
%
%
\subsubsection{Mean~$F_{22}^{\zeta_v}$ and ensemble averaged solution}
\label{Section.MeanF22}
\begin{figure}
	\centering
	\mbox{}\hspace{2.0em}\includegraphics[scale=1]{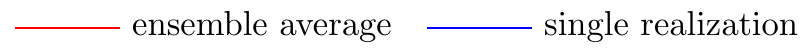}\vspace{-0.5em}
	\subfloat[$L/\ell = 80$]{\includegraphics[scale=1]{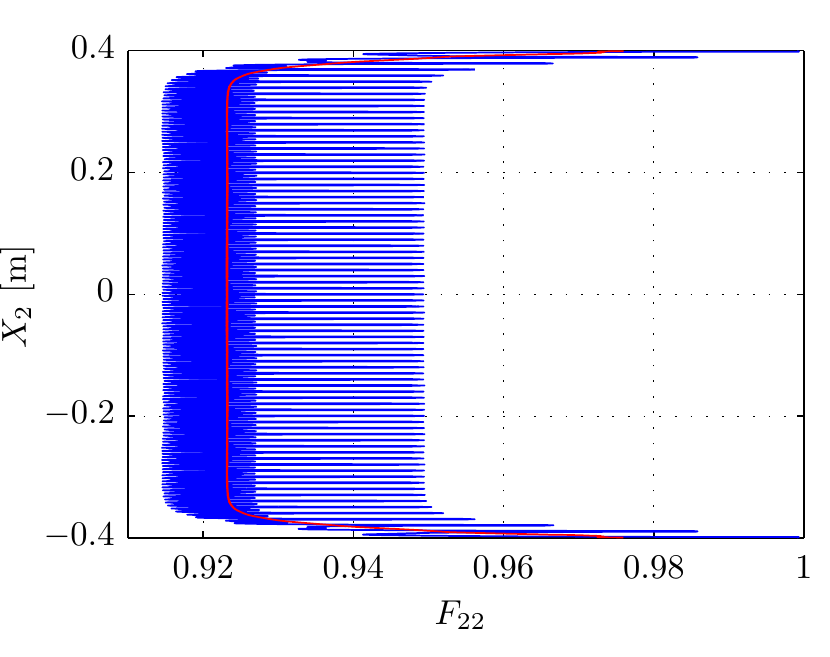}\label{Figure.meanF22a}}		
	\subfloat[$L/\ell = 20$]{\includegraphics[scale=1]{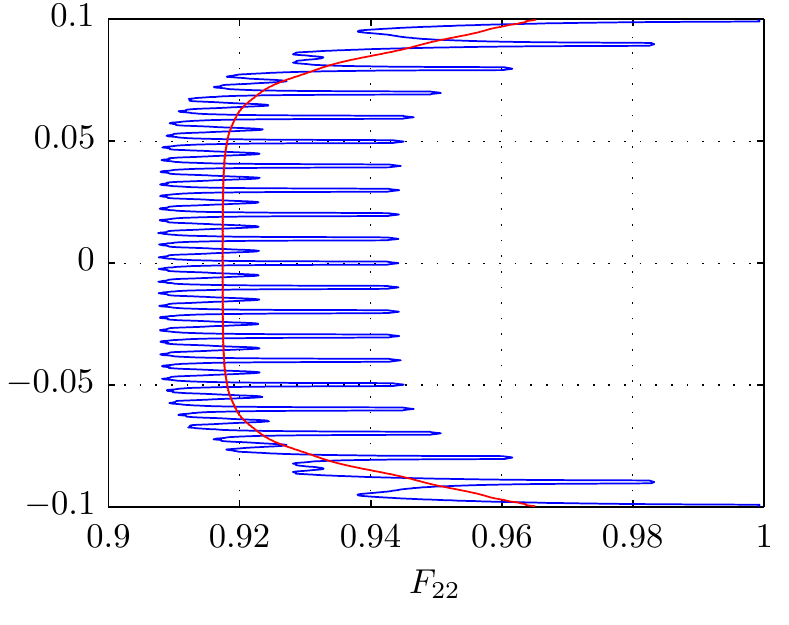}\label{Figure.meanF22b}}\\
	\subfloat[$L/\ell = 10$]{\includegraphics[scale=1]{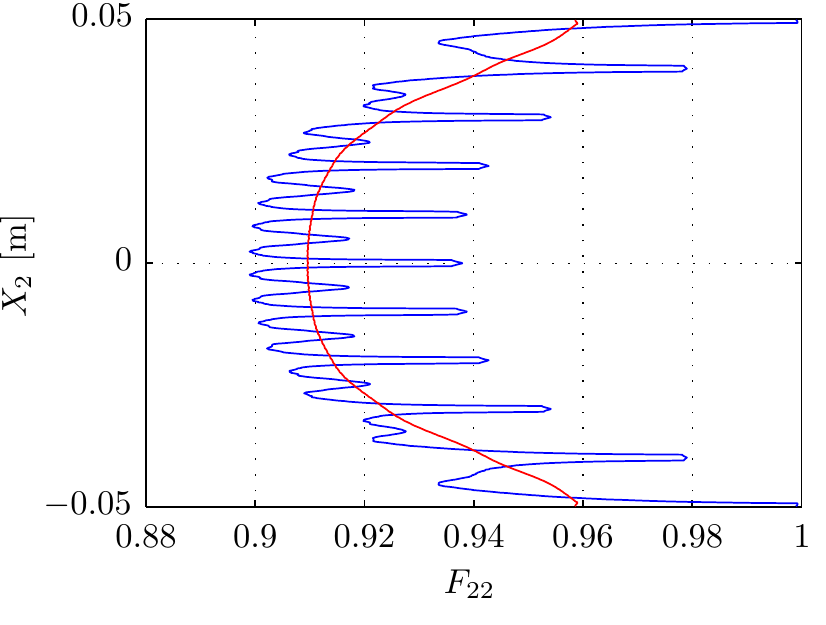}\label{Figure.meanF22c}}
	\subfloat[$L/\ell = 5$]{\includegraphics[scale=1]{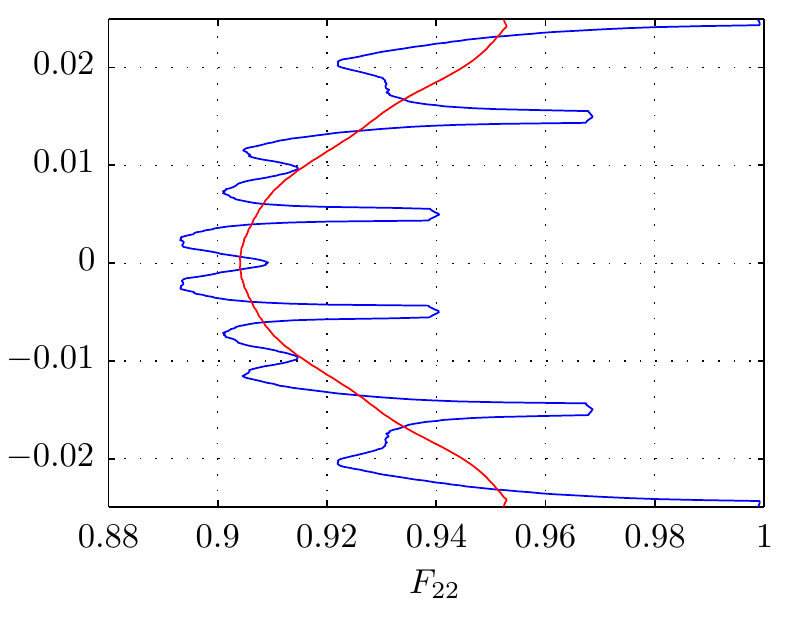}\label{Figure.meanF22d}}	
	\caption{Mean~$F_{22}^{\zeta_v}$ against~$X_2$ for the zero shift (in blue), and the ensemble averaged solution for the entire family of shifted microstructures (in red) corresponding to four scale ratios. In all cases, $u/L = 0.075$ (post-bifurcation regime).}
	\label{Figure.meanF22}
\end{figure}

For each scale ratio and all realizations, the $F_{22}$ component of the deformation gradient tensor~$ \bs{F} $ (reflecting the strain in the vertical direction) is computed on all the sampling points. These values are subsequently averaged along horizontal sections (as a numerical counterpart to Eq.~\eqref{Eq:meanP}) by taking the mean over all points lying in the corresponding horizontal section. The blue curves in Fig.~\ref{Figure.meanF22} show the mean~$F_{22}^{\zeta_v}(X_2)$ along the height direction~$\vec{e_2}$ for specimens with scale ratio~$L/\ell = 80$, $20$, $10$, $5$, and for zero vertical shifts. All curves clearly display fast-scale fluctuations due to the microstructure. The ensemble average solution, recall Eq.~\eqref{Eq:ensembleAv}, is then obtained and plotted as the red curves in the figure. These are smooth curves, as the ensemble averaging smoothens out the fast-scale fluctuations, providing the average slow-scale solution. The ensemble averaged solution indicates the region of uniform macroscopic strain, which corresponds to the plateau in the center, and the boundary layers near the two surfaces. For the scale ratio~$L/\ell = 20$, about~$70\,\%$ of the problem geometry experiences uniform macroscopic strain. For~$L/\ell \leq 7$, boundary layers dominate.
%
%
\subsubsection{Quantitative comparison}
\label{QuantCompar}
\begin{figure}
	\centering
	\includegraphics[scale=1]{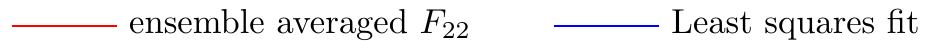}\\\vspace{-1.0em}
	\subfloat[$L/\ell = 20$]{\includegraphics[scale=1]{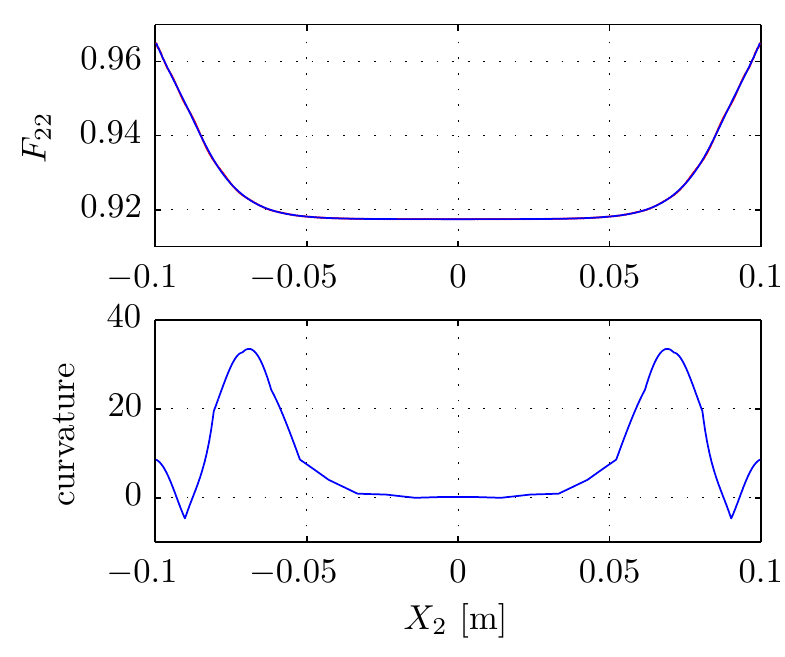}\label{Figure.bla}}\hspace{0.0em}
	\subfloat[$L/\ell = 5$]{\includegraphics[scale=1]{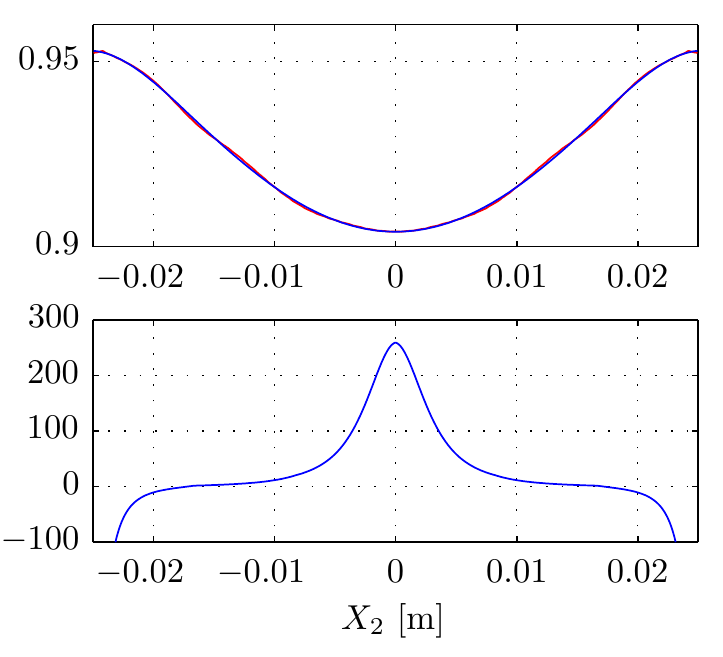}\label{Figure.blb}}
	\caption{Boundary layer thickness: the ensemble averaged~$F_{22}$ and its $C^2$-continuous least squares fit corresponding to two scale ratios: (a)~$L/\ell = 20$ and (b)~$L/\ell = 5$. In both cases, $u/L = 0.075$ (post-bifurcation regime).}
	\label{Figure.bl}
\end{figure}

\begin{figure}
	\centering
	\includegraphics[scale=1]{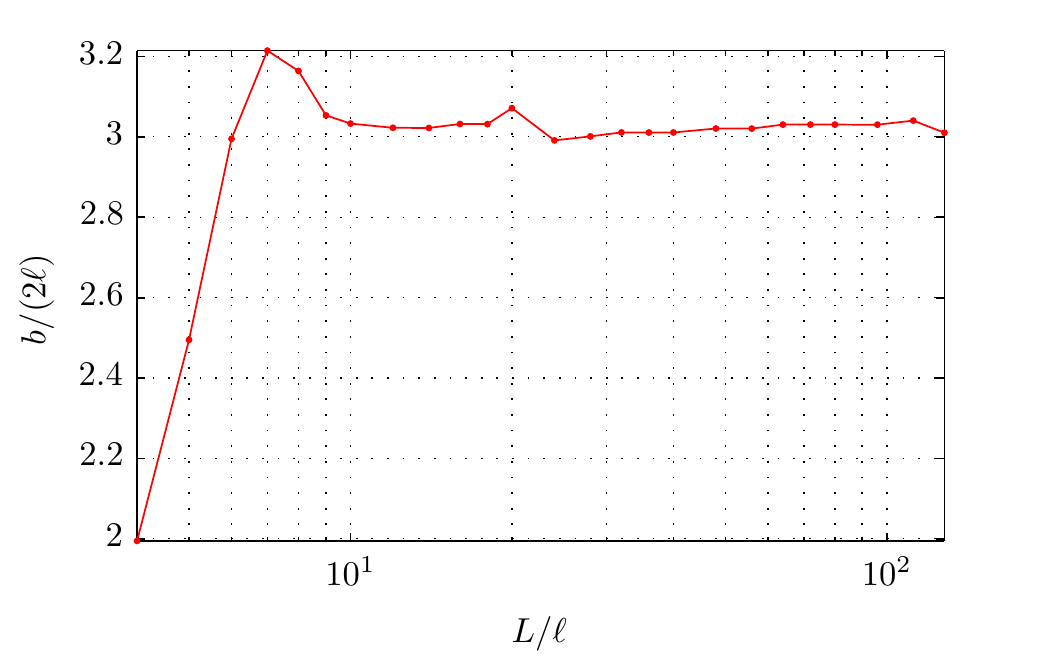}
	\caption{Number of patterned unit cells in each boundary layer, $ b/(2\ell) $, as a function of the scale ratio~$L/\ell$ for applied nominal strain~$u/L = 0.075$ (post-bifurcation regime).}
	\label{Figure.bl_vs_sr}
\end{figure}

%
The boundary layer thickness~$ b $ is next defined by making use of the points of maximum curvature of a $C^2$-continuous least squares fit of the $F_{22}$ component of the deformation gradient tensor~$ \bs{F} $, see Fig.~\ref{Figure.bl}. For specimens showing two points of maximum curvature (cf. Fig.~\ref{Figure.bla}), $ b $ is defined as the distance from each of the specimen's boundaries to the closer of the two points with the maximum curvature ($b$ is actually averaged over both instances to rectify possible deviations from ideal symmetry). For smaller specimens, which have only one point of maximum curvature at the center (cf. Fig.~8b), one half of the specimen's height~$L$ is considered as boundary layer thickness~$b$. The boundary layer thickness~$b$ is further normalized by the unit cell size~$ \ell $ in order to extract the number of patterned unit cells composing the boundary layers, and is plotted as a function of the scale ratio in Fig.~\ref{Figure.bl_vs_sr}. This result shows that the number of unit cells in each boundary layer, $ b/(2\ell) $, approaches the value of~$3$ for large scale ratios~$L/\ell$, whereas it follows a linear trend for~$L/\ell < 7$. For small scale ratios this effectively means that the boundary layer propagates throughout the entire height of the specimen~$L$, cf. also Fig.~\ref{Figure.blb}.

\begin{figure}
	\centering
	\includegraphics[scale=1]{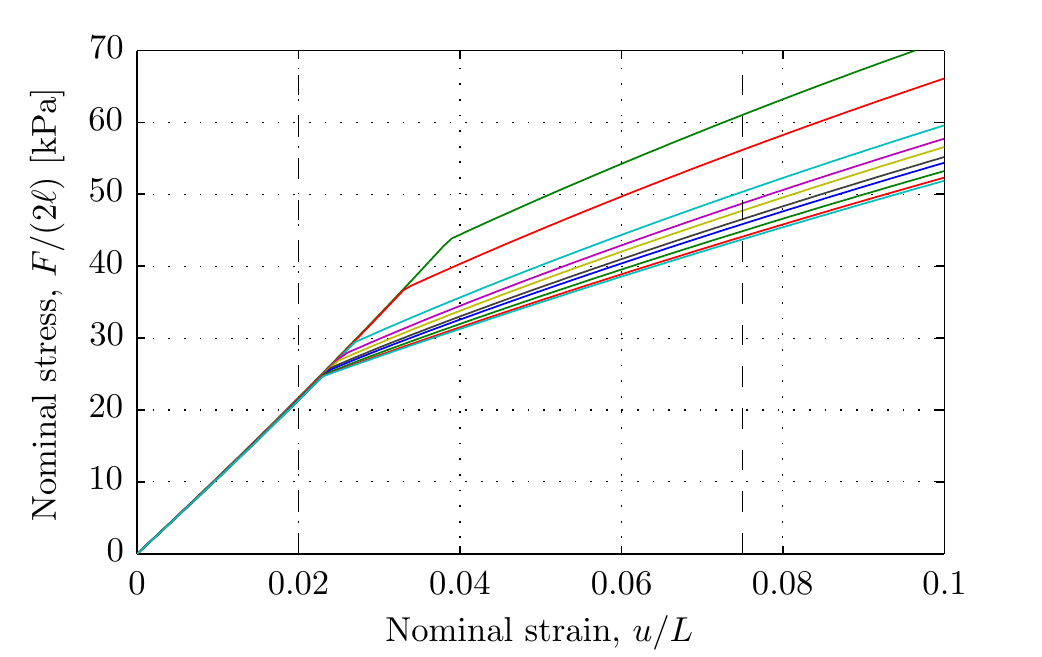}
	\includegraphics[scale=1]{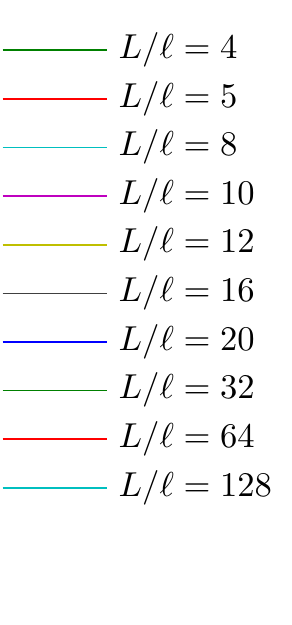}
	\caption{Stress--strain curves corresponding to the realization~$\zeta_v = 0$ and various scale ratios. The two dashed vertical lines indicate the levels of nominal strains at which the scale separation curves are plotted in Figures~\ref{Figure.ssc} and~\ref{Figure.skc}.}
	\label{Figure.stress-strain}
\end{figure}

The next quantity we focus on is the nominal stress~$ F/(2\ell) $, shown in Fig.~\ref{Figure.stress-strain} for ten scale ratios and zero vertical shift. The figure shows that all curves exhibit approximately bilinear behavior with almost identical initial linear parts and collinear responses in the post-bifurcation regimes. Because ideal geometries have been used, all bifurcation points are distinguished by sharp kinks in the corresponding stress--strain curves, whose positioning strongly depends on the scale ratio. Overall, the specimens with more unit cells (a higher scale ratio~$L/\ell$) bifurcate at a smaller strain compared to the cases with fewer holes.

\begin{figure}
	\centering
	\includegraphics[scale=1]{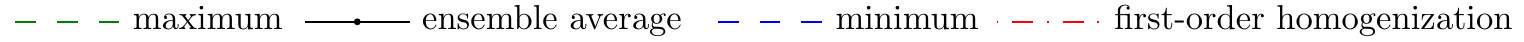}\vspace{-1.0em}\\
	\subfloat[$u/L = 0.02$]{\includegraphics[scale=1]{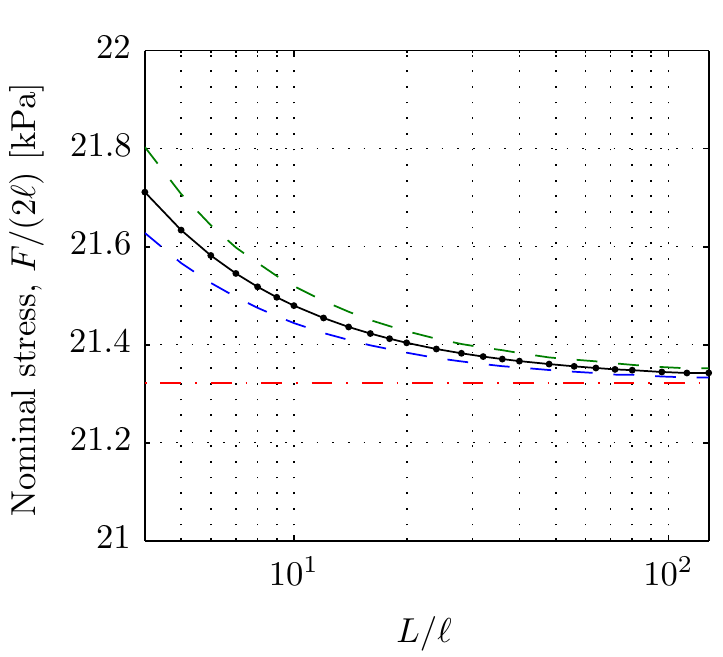}\label{Figure.ssca}}\hspace{1.0em}
	\subfloat[$u/L = 0.075$]{\includegraphics[scale=1]{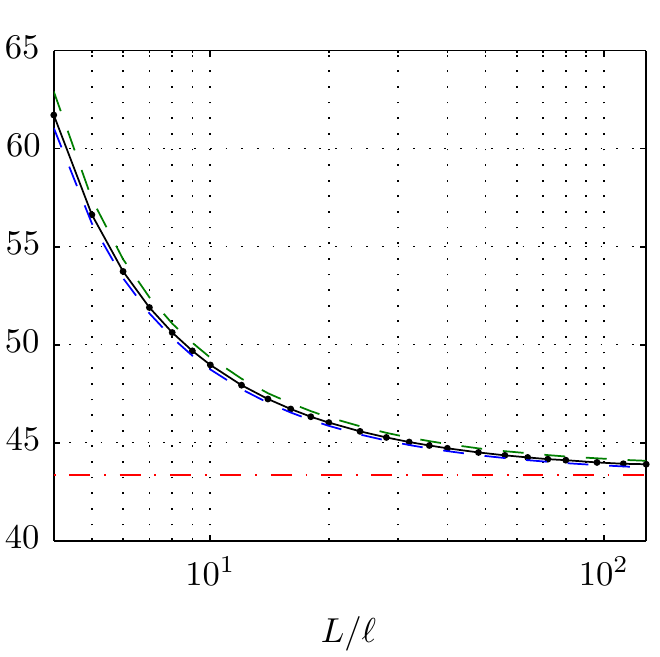}\label{Figure.sscb}}	
	\caption{Scale separation curves for the compression case showing the ensemble averaged nominal stress~$ F/(2\ell) $ as a function of the scale ratio~$L/\ell$, its maximum and minimum among all shifted realizations, and the first-order computational homogenization solution for applied nominal strain (a)~$ u/L = 0.05 $ (linear regime), and~(b) $ u/L = 0.075 $ (post-bifurcation regime).}
	\label{Figure.ssc}
\end{figure}

In order to extract behavior of nominal stress as a function of the scale ratio, two arbitrary strain levels before and beyond the bifurcation points are considered and compared, namely~$u/L = 0.02$ (early linear regime) and~$u/L = 0.075$ (deep post-bifurcation regime), cf. Fig.~\ref{Figure.stress-strain}. The result is shown in Fig.~\ref{Figure.ssc}, where the ensemble averaged nominal stress~$ F/(2\ell) $ at both strain levels is plotted against the scale ratio~$L/\ell$ with solid black lines with dots indicating the scale ratios at which the (full-scale) simulations are carried out. The individual solutions for different realizations of the microstructure fluctuate around this average solution. In order to quantify the magnitude of this fluctuation, the maximum and minimum among all realizations for each scale ratio are interconnected, spanning the two dashed curves. 

The presented results show that whereas the nominal stress observed in the linear regime changes only mildly with the scale ratio (corresponding change is less than~$2\,\%$), a strong dependence can be observed in the post-bifurcation regime (where the effect achieves approximately~$42\,\%$).

For large scale ratios, i.e.~$L/\ell \geq 50$, the ensemble averaged stress approaches a constant value of about~$ 44 $~kPa (for~$u/L = 0.075$). This is in accordance with the expectation that in the limit~$L/\ell \rightarrow \infty$, the material effectively behaves as a homogeneous medium as the microstructure becomes negligibly fine with respect to the structural length scale. This is also considered as the scale separation limit implicitly adopted in a scale-independent classical first-order computational homogenization method, recalled for the reader's convenience in~\ref{Section:A}. For constructing the first-order computational homogenization solution, the size of representative volume element (RVE) is chosen to be~$2 \times 2$ unit cells,  based on prior insights from full scale simulations performed here, and also corroborated by experimental evidence (see e.g. \citealp{Bertoldi2008d}). Accordingly, the first-order computational homogenization solutions, plotted as the red dash-dot lines in Fig.~\ref{Figure.ssc}, clearly show a good match with the ensemble averaged solutions for scale ratios~$L/\ell \geq 50$, for which the observed deviations are less than~$3\,\%$. Solutions for larger scale ratios therefore behave as if the size of the microstructural features is insignificant, not affecting the ensemble averaged stress.
\begin{figure}
	\centering
	\includegraphics[scale=1]{Fig_11_legend.pdf}\vspace{-1.0em}\\
	\subfloat[$u/L = 0.02$]{\includegraphics[scale=1]{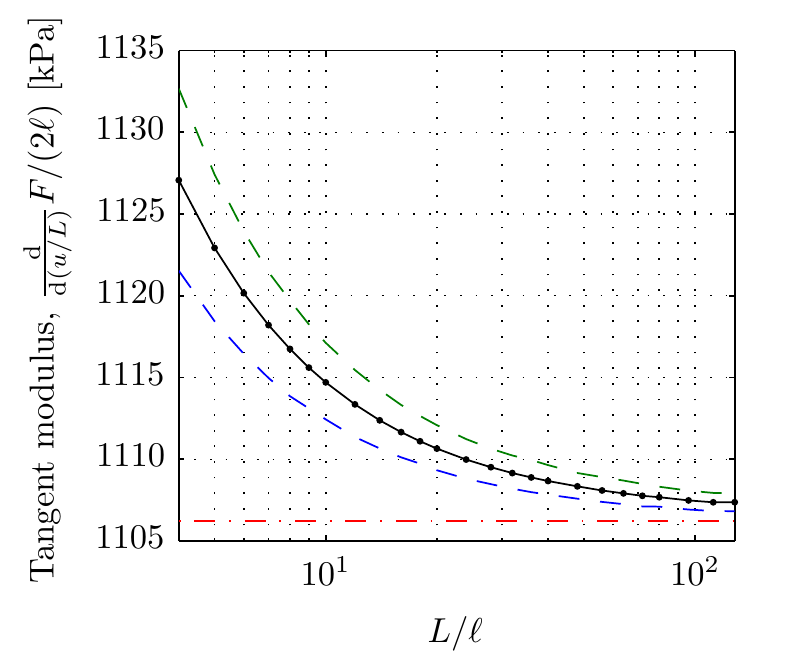}\label{Figure.skca}}\hspace{1.0em}
	\subfloat[$u/L = 0.075$]{\includegraphics[scale=1]{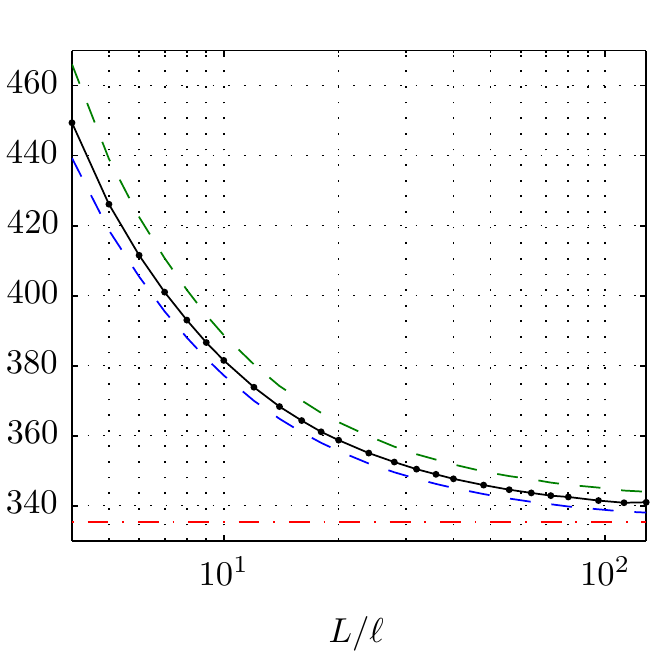}\label{Figure.skcb}}	
	\caption{Scale separation curves for the compression case showing the ensemble averaged tangent modulus as a function of the scale ratio~$L/\ell$, its maximum and minimum among all realizations, and the first-order computational homogenization solution for applied nominal strain (a)~$ u/L = 0.05 $ (linear regime), and~(b) $ u/L = 0.075 $ (post-bifurcation regime).}
	\label{Figure.skc}
\end{figure}

For scale ratios~$L/\ell \leq 50$, a non-constant stress~$ F/(2\ell) $ emerges, i.e. the solution becomes scale-dependent. The microstructural features are now large enough, relative to the structural length scale, to influence the mean solution. The averaged solution increases and reaches its maximum at the lowest scale ratio considered, $L/\ell = 4$. At this point, the mean stress is about~$ 42\,\% $ higher than its corresponding value within the scale-independent limit. This deviation of the mean solution cannot be captured using the scale-independent first-order computational homogenization method.

The bounding curves of the fluctuations in Fig.~\ref{Figure.ssc}, representing the maximum and minimum fluctuation of individual realizations, show that the fluctuations are higher for smaller scale ratios. Note that these extremes correspond to configurations with approximately~$\zeta_v = 0$ (minimum) and~$\zeta_v = 0.5$ (maximum). For scale ratios~$L/\ell \geq 50$, the peak-to-peak amplitude of fluctuation reduces to a constant value of about~$ 0.4$~kPa (for~$u/L = 0.075$). This is less than~$ 1\,\% $ of the constant mean nominal stress of~$ 44 $~kPa. This means that for~$ L/\ell \geq 50 $, the ensemble averaged solution is very close to the solution of individual realizations, which also substantiates fewer number of shifts used. For smaller scale ratios, however, the fluctuations are higher, but the maximum bandwidth of fluctuations at a scale ratio~$ L/\ell = 4 $ remains less than~$ 5\,\% $, compared to the corresponding deviation of the mean value which exceeds~$ 40\,\% $.

It is interesting to note that when the stress--strain curve of Fig.~\ref{Figure.sscb} is plotted in the log-log scaling, an affine function with slope approximately~$-1$ results. Such a behavior can be accurately captured by a simplified continuous sandwich model when the thickness of the boundary layer is considered as a known parameter, recall Fig.~\ref{Figure.bl_vs_sr}.

Fig.~\ref{Figure.skc} shows the scale separation curves representing the tangent modulus for the compression case, which is defined as {\small $\frac{\mathrm{d}}{\mathrm{d}(u/L)}F/(2\ell)$}, and for the two fixed levels of applied nominal strain. These curves display a very similar qualitative as well as quantitative behavior to those of the nominal stress. A change in the tangent modulus corresponding to the two nominal strains~$0.02$ and~$0.075$ is approximately~$1/3$ due to pattern transformation.
%
%
\subsection{Bending}
\label{Bending}
Before presenting the results for the bending case, the normalization employed throughout this section is first clarified. Graphs of nominal stress and strain will be presented, which follow from the standard theory of Bernoulli beams. As a rotation angle~$\theta$ is prescribed to the modeled specimen (recall Fig.~\ref{Figure.Sketch2}(b)), the corresponding strain at a given vertical coordinate~$X_2$ reads as
\begin{equation}
\varepsilon(X_2) = \kappa X_2, \quad \frac{-L}{2} \leq X_2 \leq \frac{L}{2},
\label{Eq:curvature}
\end{equation}
where~$\kappa = \theta/ (2 \ell)$ is the curvature. The ``nominal" strain refers to the values at the bottom or top edge, i.e.~$\varepsilon(\pm L/2) = \pm \theta L/(4 \ell)$. For all simulations, the rotation angle~$\theta$ is prescribed such that the bottom edge is subjected to a maximum nominal compressive strain~$\theta L/(4 \ell) = 0.15 $, i.e.~$50\,\%$ higher than the maximum uniform strain applied in the case of uniaxial compression. As for the normal stress, one has
\begin{equation}
\sigma(X_2) = \frac{MX_2}{I_\mathrm{X_3}} + \frac{N}{A}, \quad \frac{-L}{2} \leq X_2 \leq \frac{L}{2},
\label{Eq:stress}
\end{equation}
where~$M$ is the bending moment, $I_{X_3}$ is the second moment of area, $N$ the reaction normal force, and~$A$ the cross-sectional area. Assuming a rectangular cross-section (of a unit width and height~$L$ due to the plane strain condition), the top and bottom edges experience a stress~$\sigma(\pm L/2) = \pm 6M/L^2 + N/L$. ``Nominal" stresses refer to the values~$6M/L^2$ (bending), or~$N/L$ (normal load), respectively.

\begin{figure}[p]
	\centering
	\subfloat[deformed configurations]{\includegraphics[scale=1]{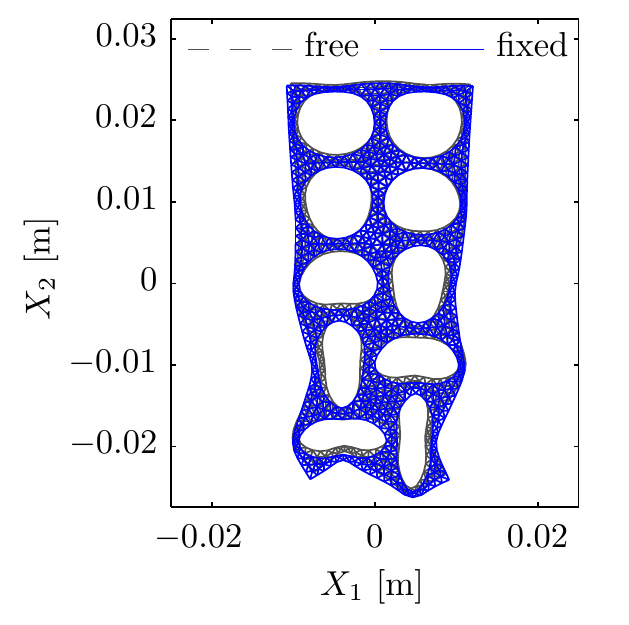}\label{Figure.bfixedvsfreea}}
	\subfloat[nominal stress--strain diagram]{\includegraphics[scale=1]{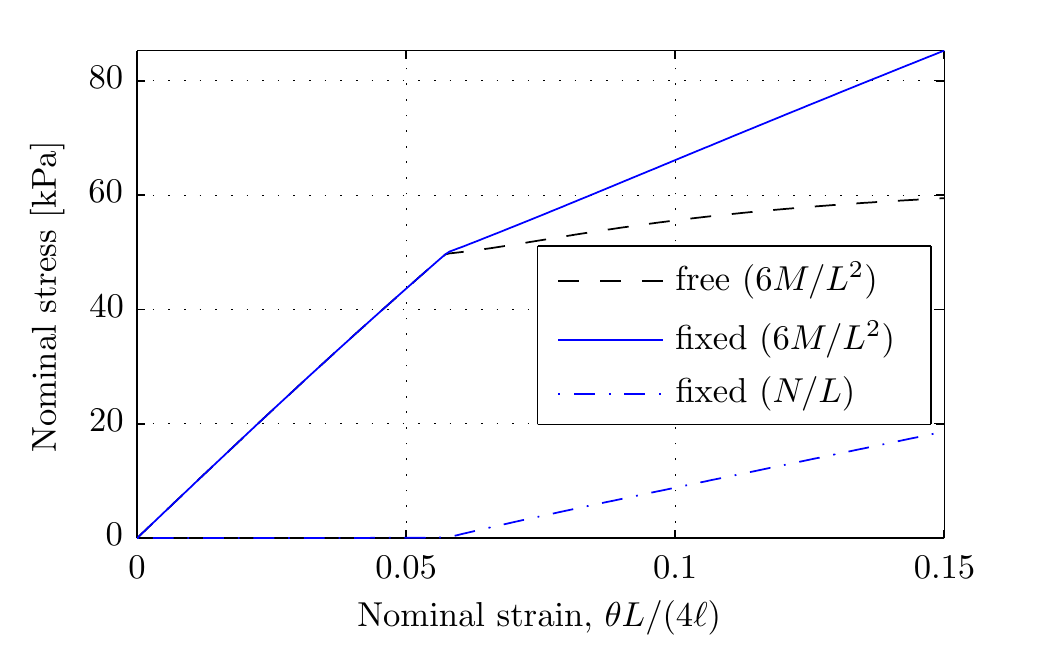}\label{Figure.bfixedvsfreeb}}	 
	\caption{Comparison between horizontally free (case~(i)) vs fixed (case~(ii)) bending for a specimen with scale ratio~$L/\ell = 5$ and~$\zeta_v = 0$. The deformed shapes are shown on the left, whereas the nominal stresses due to bending~($6M/L^2$) and tension~($N/L$) vs the applied nominal strain~($\theta L/(4\ell)$) are depicted on the right.}
	\label{Figure.bfixedvsfree}
\end{figure}

\begin{figure}[p]
	\centering
	\hspace{3em}\includegraphics[scale=1]{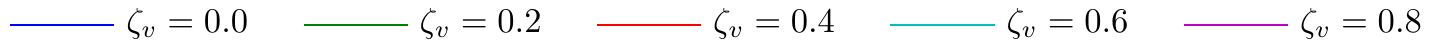}\vspace{-0.75em}\\
	\subfloat[free]{\includegraphics[scale=1]{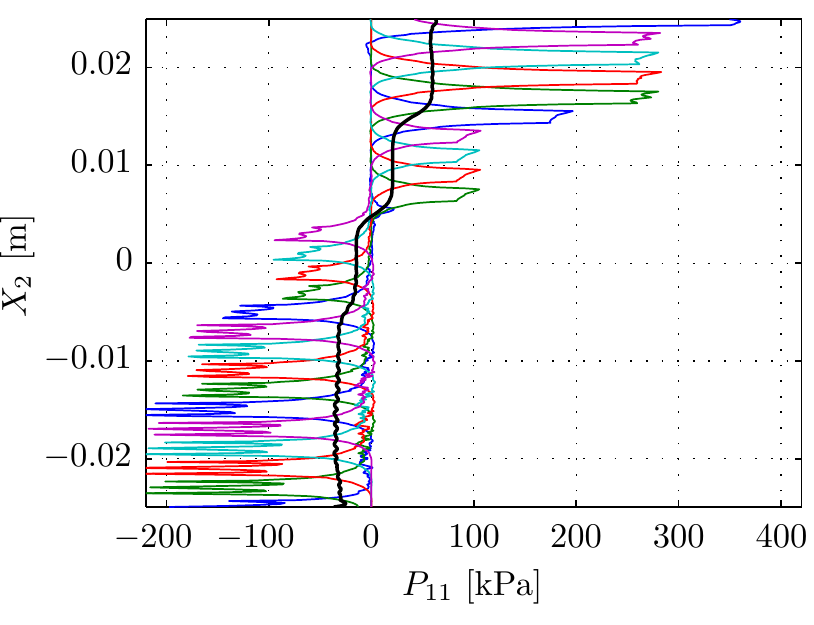}\label{Figure.bshifteda}}\hspace{0.0em}
	\subfloat[fixed]{\includegraphics[scale=1]{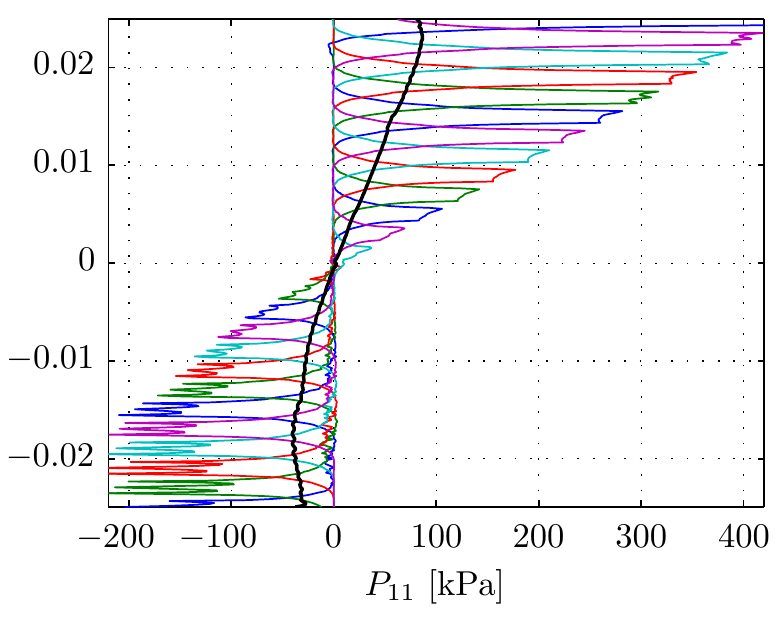}\label{Figure.bshiftedb}}	 
	\caption{Mean stress~$P_{11}^{\zeta_v}$ along~$X_2$ for five vertical shifts~$\zeta_v$ and scale ratio~$L/\ell = 5$. (a)~Horizontally unconstrained bending (case~(i)), and~(b) horizontally constrained bending (case~(ii)). The ensemble averaged solutions are shown as thick black lines. All results correspond to~$\theta L/(4\ell) = 0.1125$ (post-bifurcation regime).}
	\label{Figure.bshifted}
\end{figure}

To justify the boundary conditions used in Section~\ref{BC}, a comparative study is made of two types of boundary conditions, complementing the prescribed rotation angle~$\theta$ and periodic boundary fluctuations:
\begin{itemize}

	\item[(i)] zero vertical displacements at points~$(\pm \ell, 0)$ and zero horizontal displacement of the left point~$(- \ell, 0)$, inducing pure bending only
	
	\item[(ii)] prescribed zero displacements at points~$(\pm \ell, 0)$ in both vertical as well as horizontal directions, which effectively results in bending combined with normal force.
	
\end{itemize}
Both cases are compared in terms of their deformed configurations and nominal stress--strain diagrams in Fig.~\ref{Figure.bfixedvsfree}, and in terms of the mean stress~$P_{11}^{\zeta_v}$ (and the ensemble averaged solutions) in Fig.~\ref{Figure.bshifted}, for the scale ratio~$L/\ell = 5$. Note that in the case of pure bending the neutral axis shifts in the vertical direction, and that the corresponding stress solutions are practically rigidly-shifted copies of each other (Fig.~\ref{Figure.bshifteda}). This shift introduces steps in the ensemble averaged solution (shown as the continuous black line), suggesting that type~(i) boundary conditions are not appropriate for investigating size effects and boundary layers. On the contrary, the presence of the horizontal constraint introduced in the second case results in a tensile normal force (cf. the blue dash-dot line in Fig.~\ref{Figure.bfixedvsfreeb}) that fixes the vertical positioning of the neutral axis at~$X_2 = 0$. The mean stress~$P_{11}^{\zeta_v}$ is now shifted \emph{and scaled} for individual configurations~$\zeta_v$. Ensemble averaging effectively eliminates all fine-scale fluctuations, resulting in the black continuous line in Fig.~\ref{Figure.bshiftedb} where no fast scale fluctuations can be observed. In some of the results presented below, small oscillations will appear again. These are, nevertheless, a result of the finite number of shifts used, as their magnitude is vanishingly small and increases with increasing scale ratio for which fewer shifts are available (cf. Fig.~\ref{Figure.stressP11a} for instance).
%
%
\subsubsection{Full-scale numerical solutions}
\label{DNSsol}
\begin{figure}
	\centering
	\subfloat[$\zeta_v = 0$]{\includegraphics[scale=0.5]{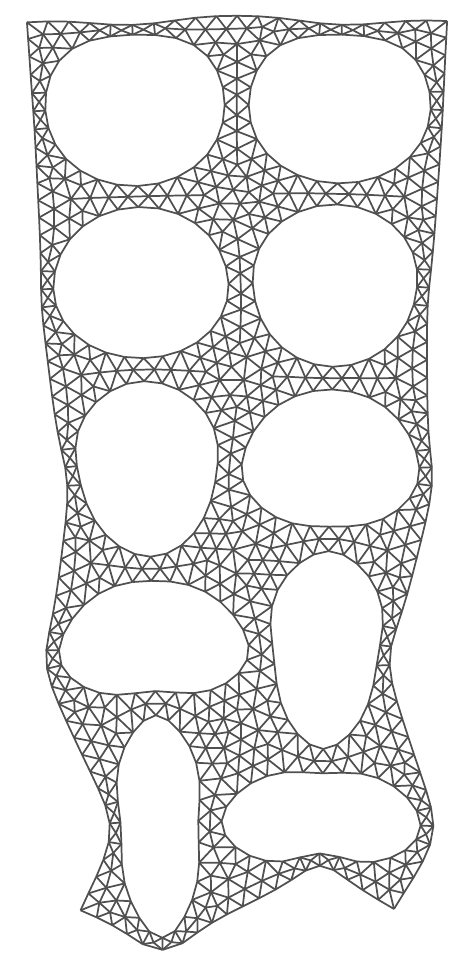}\label{Figure.shifts_bendinga}}\hspace{1.0em}
	\subfloat[$\zeta_v = 0.25$]{\includegraphics[scale=0.5]{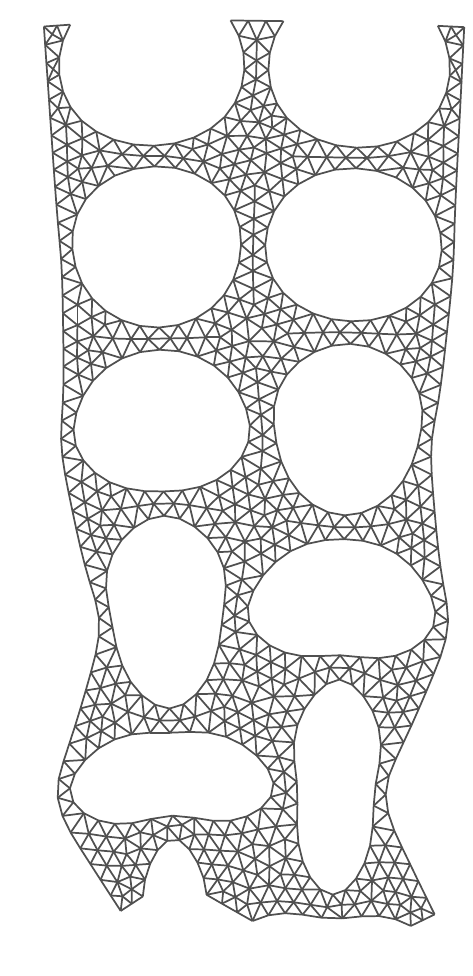}\label{Figure.shifts_bendingb}}\hspace{1.0em}
	\subfloat[$\zeta_v = 0.5$]{\includegraphics[scale=0.5]{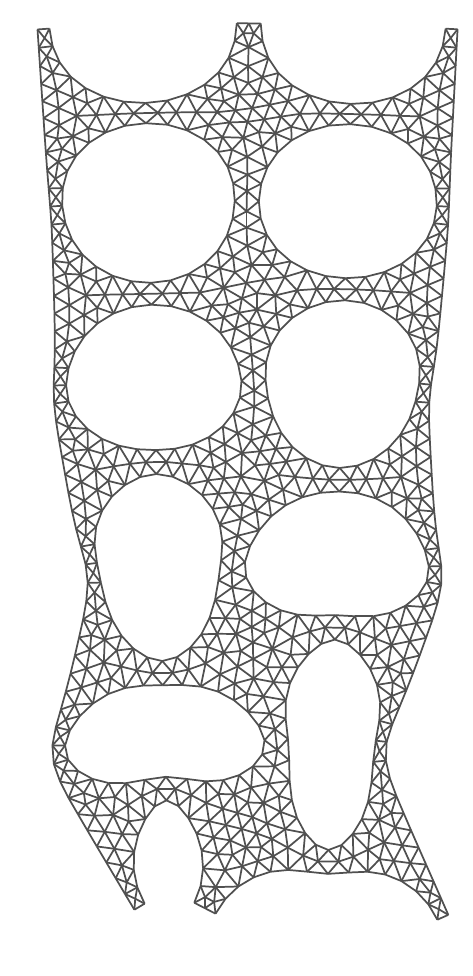}\label{Figure.shifts_bendingc}}\\
	\subfloat[stress--strain diagram]{\includegraphics[scale=1]{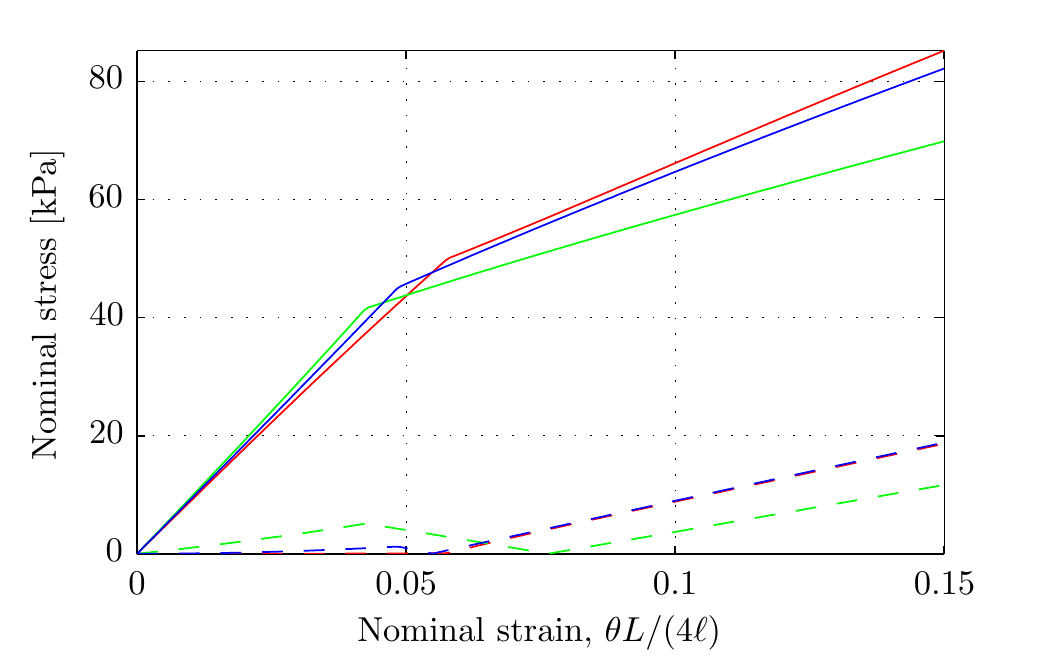}\includegraphics[scale=1]{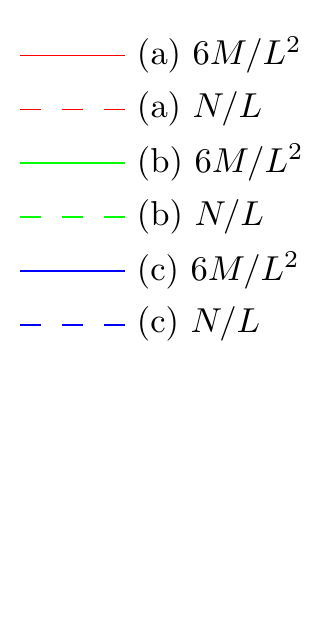}\label{Figure.shifts_bendingd}}	
	\caption{Deformed shapes under bending for the scale ratio~$ L/\ell = 5 $ and three vertical shifts~$ \zeta_v $ of the microstructure: (a)~$ \zeta_v = 0 $, (b)~$ \zeta_v = 0.25 $, and~(c) $ \zeta_v = 0.5 $. The stress--strain curves for all three cases are shown in~(d), with contributions from the reaction moment, $6M/L^2$, and normal force, $N/L$.}
	\label{Figure.shifts_bending}
\end{figure}

Fig.~\ref{Figure.shifts_bending} shows three deformed configurations (corresponding to the specimens shown in Fig.~\ref{Figure.shifts}) for the scale ratio~$L/\ell = 5$ and corresponding nominal stress--strain diagrams. The deformed configurations indicate the influence of the vertical shift~$ \zeta_v $ of the microstructure in the specimen. The effect of the free top/bottom boundaries, as opposed to the fixed boundaries used in the compression case, can also be observed. The gradient in the macroscopic strain along the height direction results in a progressive pattern transformation occurring in the lower part of the specimen, i.e. below the neutral axis. 

As opposed to the compression case, differences between individual nominal stress--strain curves are more pronounced (see Fig.~\ref{Figure.shifts_bendingd}), i.e. the vertical positioning of the microstructure~$\zeta_v$ has a larger influence. Although all curves display approximately bilinear behavior, the slopes differ significantly, even in the linear regime. This is a direct consequence of the fact that when holes are intersecting the top and bottom boundaries, no normal stress can be transmitted through these regions (approximately of size~$\ell$). Moreover, these edges are far from the neutral axis, resulting in a significant contribution to the nominal stress and stiffness. This effect is expected to vanish for increasing scale ratio~$L/\ell$ as the microscopic length~$\ell$ becomes negligible relative to the height of the specimen~$L$. The contribution of normal forces to the nominal stress in these specimens (dashed curves in Fig.~\ref{Figure.shifts_bendingd}) is significantly lower compared to that of the reaction moment. Note that some of these curves show a negative post bifurcation stiffness (cf. e.g. the green dashed curve corresponding to $\zeta_v = 0.25$). This is due to local instabilities and the auxetic nature of the considered microstructure. The global response, however, remains stable throughout the entire loading path. Furthermore, the bifurcation points, at which patterning initiates, cluster around a nominal strain of approximately~$0.05$, and a nominal stress of $50$~kPa. This is in close agreement with the values obtained for the uniform compression case, see Fig.~\ref{Figure.fixedvsfreeb}.
%
%
\subsubsection{Influence of the scale ratio~$L/\ell$}
\label{InflScaleRatio}
\begin{figure}
	\centering
	\subfloat[$L/\ell = 20$]{\includegraphics[scale=0.7,angle =90]{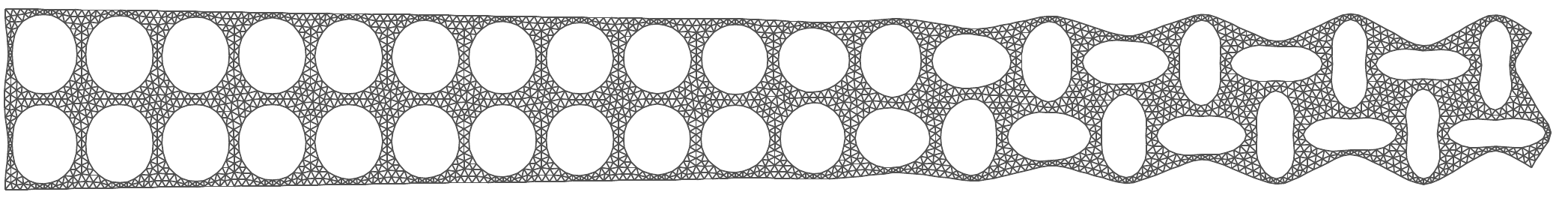}\label{Figure.sr_benda}}\\
	\subfloat[$L/\ell = 10$]{\includegraphics[scale=0.7,angle =90]{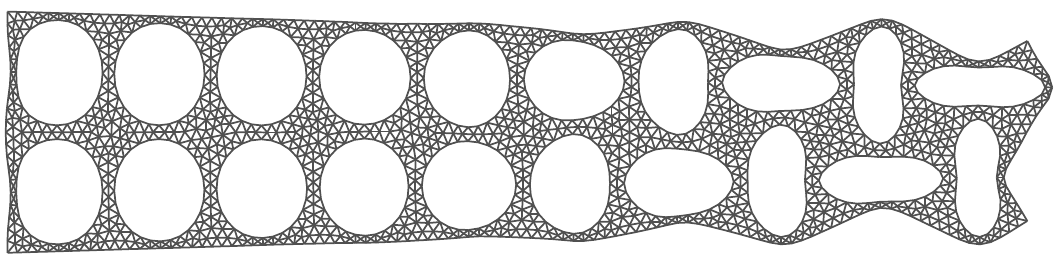}\label{Figure.sr_bendb}}\hspace{2.0em}
	\subfloat[$L/\ell = 5$]{\includegraphics[scale=0.7,angle =90]{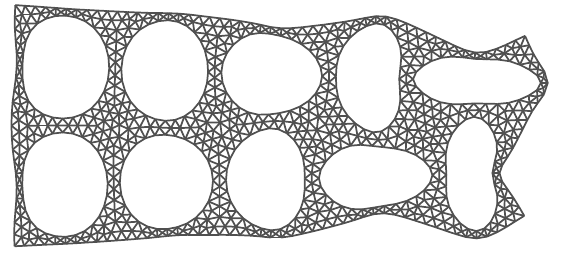}\label{Figure.sr_bendc}}	
	\caption{Deformed shapes under bending (rotated~$90$\textdegree~anticlockwise) for specimens with zero shifts in microstructure and for three scale ratios.}
	\label{Figure.sr_bend}
\end{figure}
In Fig.~\ref{Figure.sr_bend}, the deformed shapes for three different scale ratios and zero vertical shift in the microstructure are shown. For the large scale ratio~$L/\ell = 20$, the effect of the strain gradient is clearly visible, and a similar pattern to the one observed in the compression case develops below the neutral axis (recall Fig.~\ref{Figure.sr_comp}). For small scale ratios, especially for~$L/\ell = 5$ shown in Fig.~\ref{Figure.sr_bendc}, a distorted shape results, as the strain gradient is too strong compared to the microstructural length~$\ell$.
%
%
\subsubsection{Mean stress~$P_{11}^{\zeta_v}$ and ensemble averaged solution}
\label{meanP11}
\begin{figure}
	\centering
	\mbox{}\hspace{2.0em}\includegraphics[scale=1]{Fig_7_legend.pdf}\vspace{-0.5em}
	\subfloat[$L/\ell = 80$]{\includegraphics[scale=1]{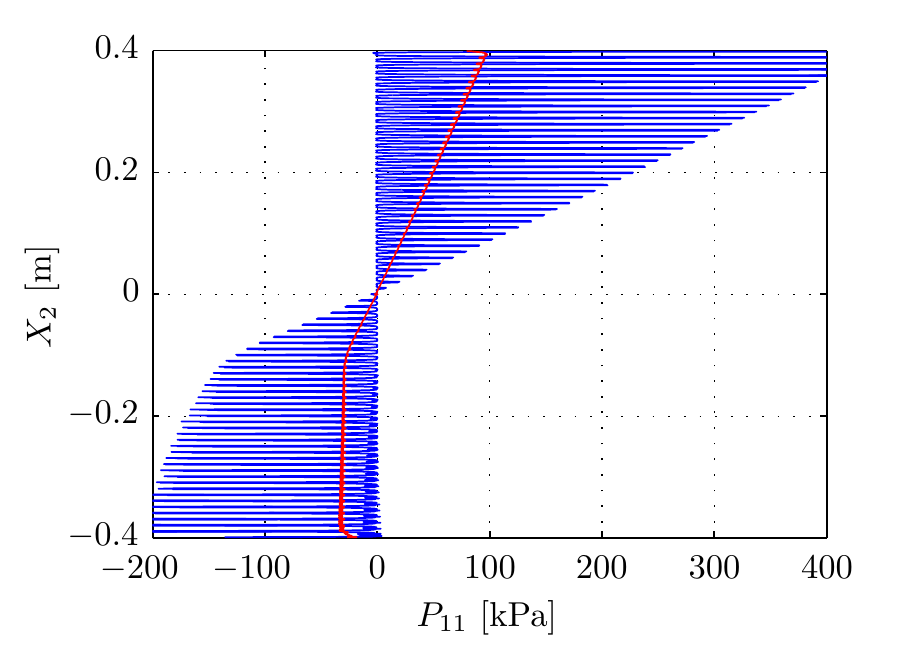}\label{Figure.stressP11a}}	
	\subfloat[$L/\ell = 20$]{\includegraphics[scale=1]{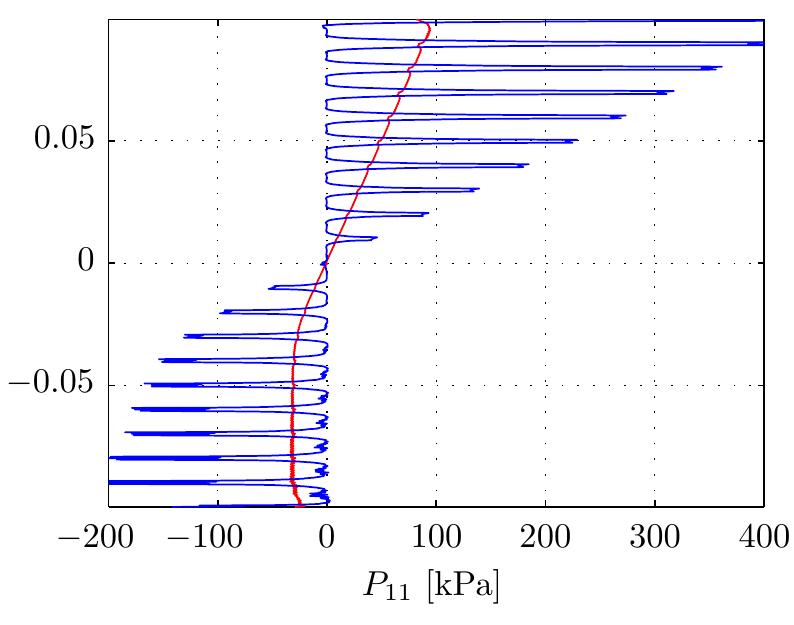}\label{Figure.stressP11b}}\\
	\subfloat[$L/\ell = 10$]{\includegraphics[scale=1]{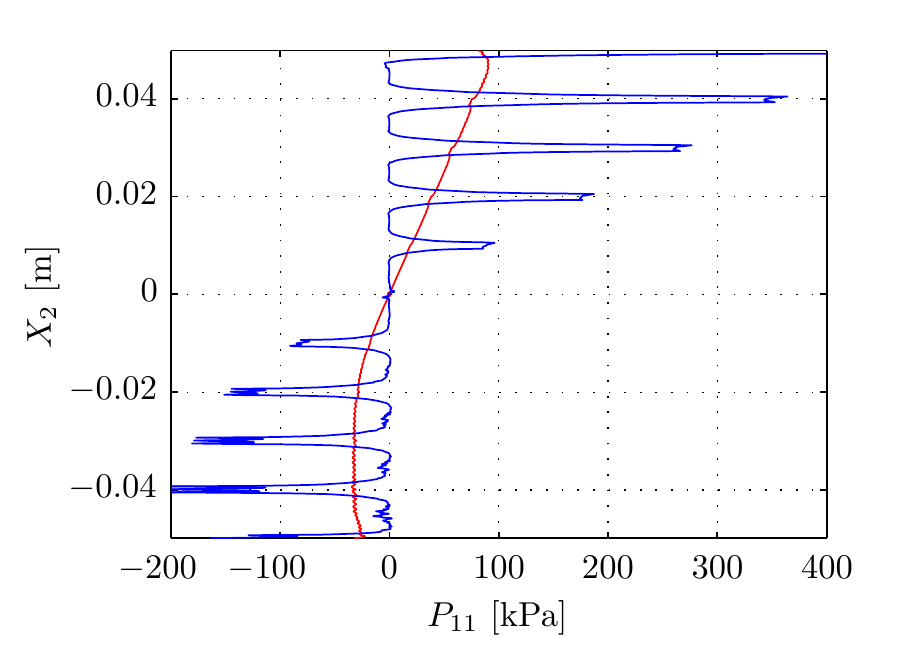}\label{Figure.stressP11c}}
	\subfloat[$L/\ell = 5$]{\includegraphics[scale=1]{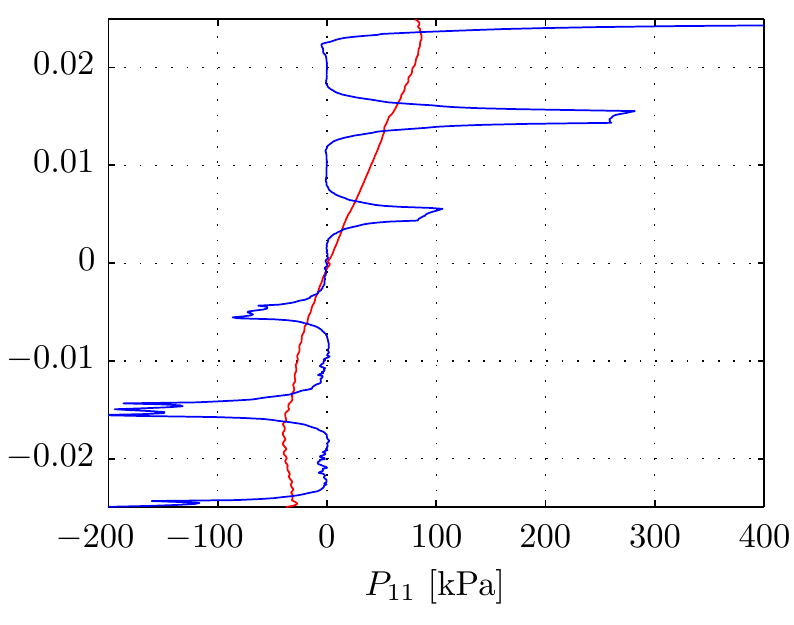}\label{Figure.stressP11d}}	
	\caption{Mean~$P_{11}^{\zeta_v}$ along~$X_2$ for the zero vertical shift (in blue), and the ensemble averaged solution for the entire family of microstructures (in red) corresponding to four scale ratios. In all cases, $\theta L/(4\ell) = 0.1125$ (post-bifurcation regime).}
	\label{Figure.stressP11}
\end{figure}

Let us now focus on the~$P_{11}$ component of the first Piola--Kirchhoff stress tensor, and its distribution among and inside specimens, and how this translates through ensemble averaging. Horizontally averaged solutions, as defined in Eq.~\eqref{Eq:meanP}, are presented in Fig.~\ref{Figure.stressP11} along the height coordinate~$ X_2 $ for specimens with scale ratios~$ L/\ell = 80$, $20$, $10$, and~$5$. The blue curves show the mean stress for the realization with zero vertical shift. In all cases, strong oscillations can be observed, peaking at the positions of the horizontal ligaments, and dropping near to zero at the spatial positions of holes. The ensemble averaged solutions, given by Eq.~\eqref{Eq:ensembleAv} and shown as the red curves in the figure, are approximately bilinear, where the linear stress profile is flattened in the bottom compressive region.

\begin{figure}[p]
	\centering
	\includegraphics[scale=1]{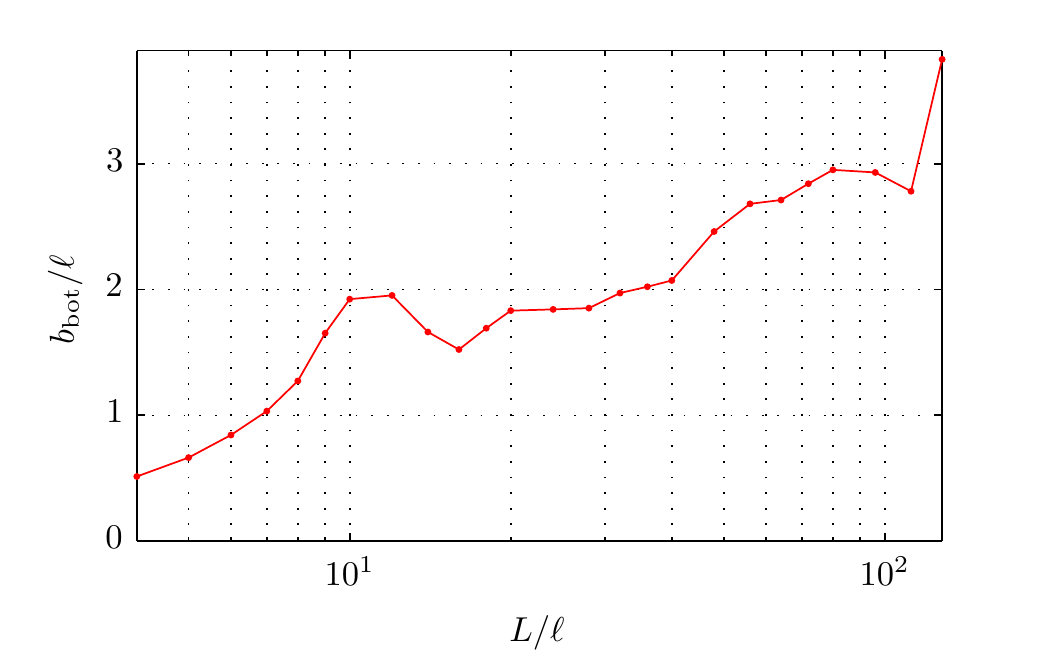}
	\caption{Number of patterned unit cells in the bottom flexible boundary layer, $ b_\mathrm{bot}/\ell $, as a function of the scale ratio~$L/\ell$ for an applied nominal strain~$\theta L/(4\ell) = 0.1125$ (post-bifurcation regime).}
	\label{Figure.bl_vs_sr_b}
\end{figure}

Close to the two free horizontal boundaries at the top and bottom parts of the model domain~$\Omega$, emerging compliant boundary layers can be observed. Since these are small, and since fewer shifts are available for higher scale ratios, the ensemble averaged solution becomes more oscillatory. As a result, the size of the bottom boundary layer normalized by the unit cell size, $b_\mathrm{low}/\ell$, is not a smooth curve in Fig.~\ref{Figure.bl_vs_sr_b}. Nonetheless, a clearly increasing trend emerges, initiating at~$0.5$ and reaching again approximately~$3$ unit cells in size. Note that the boundary layer thickness has been determined using the same method as for the compression case, i.e. through the point of the maximum curvature of a $C^2$-continuous cubic Least squares fit (recall Fig.~\ref{Figure.bl}).
%
%
\subsubsection{Quantitative comparison}
\label{QuandComp}
\begin{figure}[p]
	\centering
	\includegraphics[scale=1]{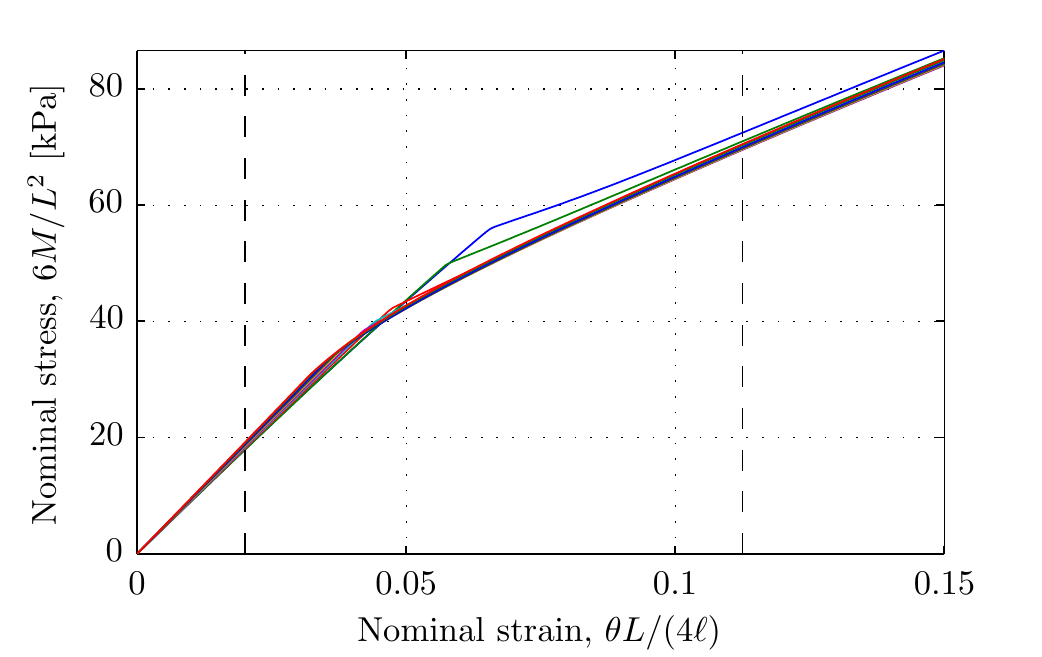}
	\includegraphics[scale=1]{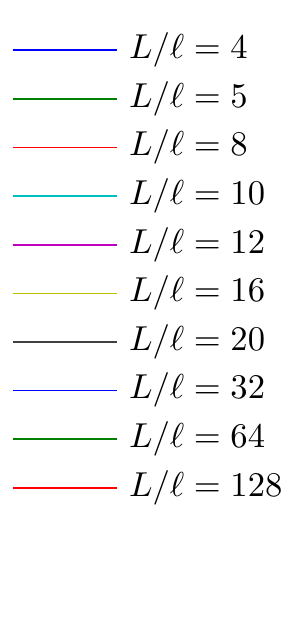}	
	\caption{Nominal stress due to the bending moment~$ 6M/L^2 $ plotted against the nominal strain~$ \theta L/(4\ell) $ for ten scale ratios~$L/\ell$, corresponding to realizations with zero vertical shifts. The dashed vertical lines indicate the levels of nominal strain at which the scale separation curves are plotted in Figures~\ref{Figure.ssc_bending} and~\ref{Figure.skc_bending}.}
	\label{Figure.moment_curves}
\end{figure}

Fig.~\ref{Figure.moment_curves} shows the nominal stress due to the bending moment plotted against the nominal strain for ten scale ratios with zero vertical shift. In contrast to the compression case, curves corresponding to various scale ratios~$L/\ell$ take different shapes. Whereas for small scale ratios a quasi-bilinear behavior is observed, for large scale ratios the transition from the linear to the bifurcated state is more gradual, resulting in smooth curves. This is a direct consequence of the presence of a strain gradient and the fact that for high scale ratios the material almost behaves as a homogeneous medium. Moreover, the initial bifurcation located closest to the bottom horizontal boundary occurs sooner for high scale ratios. At fixed nominal strain the curves with high scale ratios tend to the same nominal stress.
\begin{figure}[p]
	\centering
	\includegraphics[scale=1]{Fig_11_legend.pdf}\vspace{-1.0em}\\
	\subfloat[$\theta L/(4\ell) = 0.02$]{\includegraphics[scale=1]{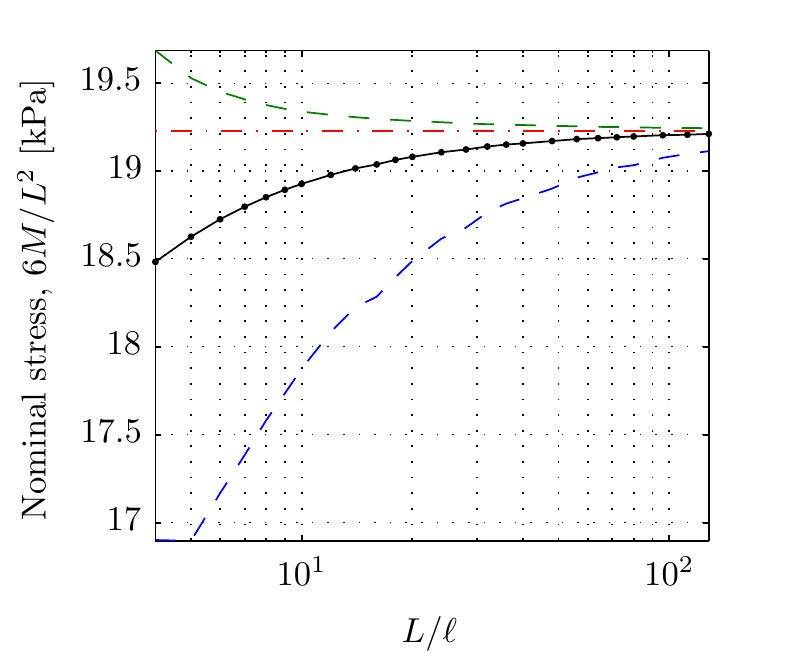}\label{Figure.ssc_bendinga}}\hspace{1.0em}
	\subfloat[$\theta L/(4\ell) = 0.1125$]{\includegraphics[scale=1]{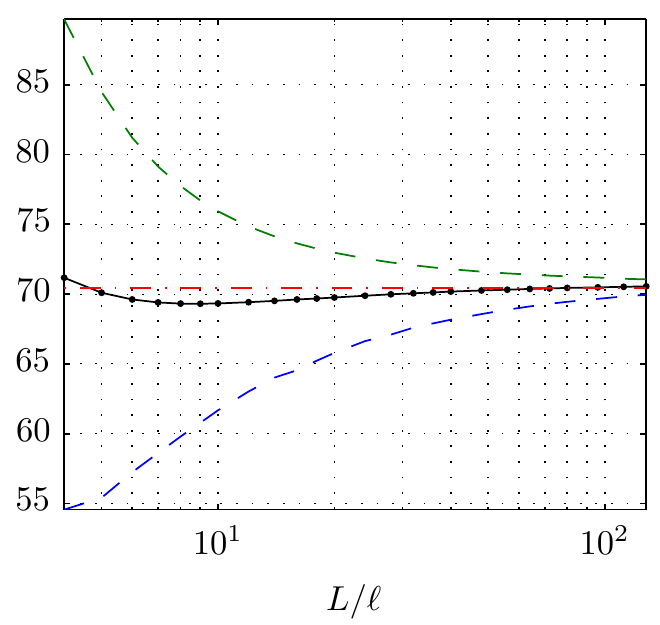}\label{Figure.ssc_bendingb}}		
	\caption{Scale separation curves for bending showing the ensemble averaged solution, maximum and minimum among all realizations, and the first-order computational homogenization solution for the nominal stress due to bending, $ 6M/L^2 $, as a function of the scale ratio $L/\ell$ at an applied nominal strain~(a) $ \theta L/(4\ell) = 0.02 $ (linear regime), and~(b) $ \theta L/(4\ell) = 0.1125 $ (post-bifurcation regime).}
	\label{Figure.ssc_bending}
\end{figure}
\begin{figure}[p]
	\centering
	\includegraphics[scale=1]{Fig_11_legend.pdf}\vspace{-1.0em}\\
	\subfloat[$\theta L/(4\ell) = 0.02$]{\includegraphics[scale=1]{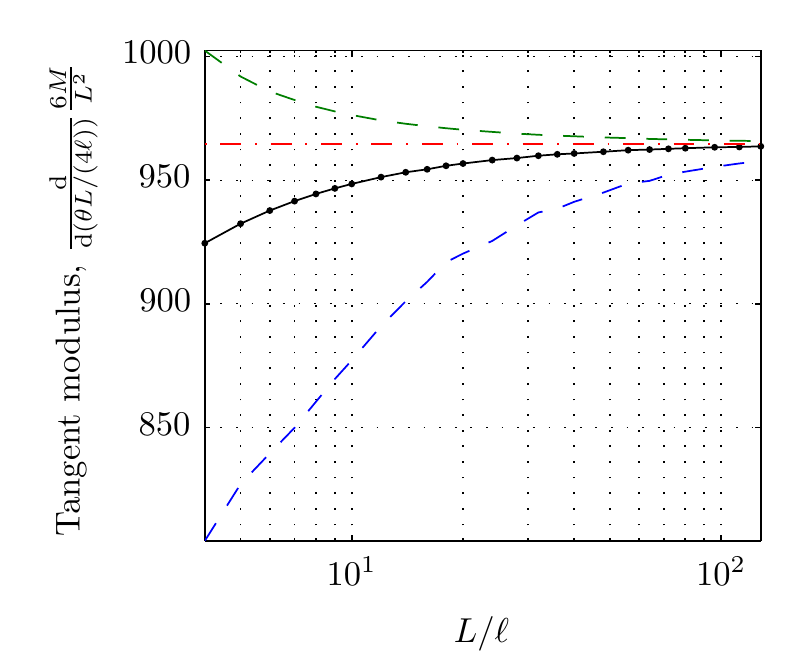}\label{Figure.skc_bendinga}}\hspace{1.0em}
	\subfloat[$\theta L/(4\ell) = 0.1125$]{\includegraphics[scale=1]{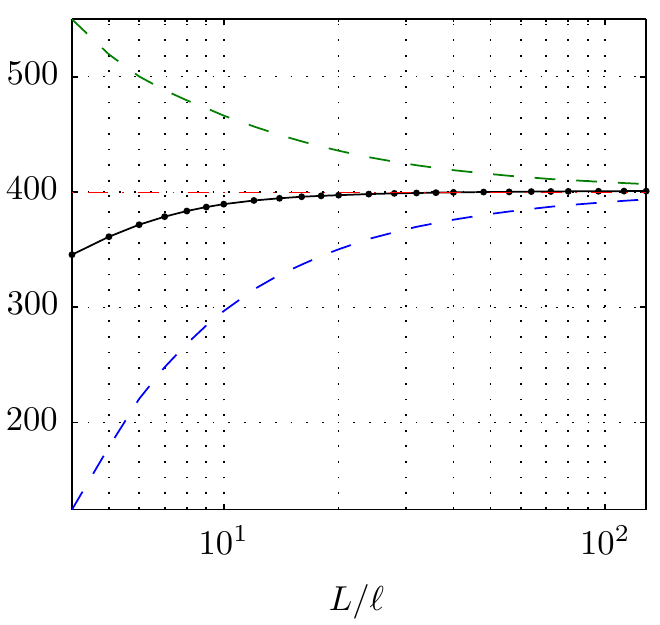}\label{Figure.skc_bendingb}}		
	\caption{Scale separation curves for bending showing the ensemble averaged tangent modulus as a function of the scale ratio~$L/\ell$, its maximum and minimum among all realizations, and the first-order computational homogenization solution for applied nominal strain~(a) $\theta L/(4\ell) = 0.02$ (linear regime), and~(b) $\theta L/(4\ell) = 0.1125$ (post-bifurcation regime).}
	\label{Figure.skc_bending}
\end{figure}

This is clearly visible in Fig.~\ref{Figure.ssc_bending}, where the ensemble averaged nominal stress due to the bending moment, $6M/L^2$, is plotted as a function of the scale ratio~$L/\ell$ for two levels of nominal strain~$\theta L/(4\ell)$. In analogy to the compression case, the strain levels are taken as~$\theta L/(4\ell) = 0.02$ (early linear regime) and~$\theta L/(4\ell) = \frac{3}{4} \cdot 0.15 = 0.1125$ (deep post-bifurcation regime), cf. Fig.~\ref{Figure.moment_curves}. The ensemble averaged solution is again shown as the solid black line, whereas fluctuations due to shifted microstructures are shown in dashed lines and the first-order homogenized solution as the red horizontal limit. The presented results reveal that for medium to large scale ratios, the ensemble averaged solutions deviate from the homogenized limits only mildly, the peak difference being less than~$4\,\%$ in both the linear as well as bifurcated regime. Naturally, extreme values occur for small scale ratios, especially for~$L/\ell \leq 10$. For higher scale ratios, the ensemble averaged stress approaches its homogenized asymptote from below, as opposed to the compression case presented in Fig.~\ref{Figure.ssc}. This behavior is a consequence of the applied boundary conditions on the top and bottom edges, i.e. fixed boundaries in the case of compression and free boundaries in the case of bending, which induce the observed stiffening and softening effects.

The dashed curves in Fig.~\ref{Figure.ssc_bending}, representing again the maximum and minimum over all realizations, reveal a significant influence of the microstructural shift~$\zeta_v$. Whereas for large scale ratios the scatter vanishes as expected, for small scale ratios it exceeds~$14\,\%$ (for~$\theta L/(4\ell) = 0.02$) and~$49\,\%$ (for~$\theta L/(4\ell) = 0.1125$) relative to the homogenized limits. The extreme values correspond to shift factors of approximately~$\zeta_v = 0.1$ (minimum) and~$\zeta_v = 0.85$ (maximum).

Finally, the tangent modulus for the bending case, defined as {\small $\frac{\mathrm{d}}{\mathrm{d}(\theta L/(4\ell))}\frac{6 M}{\ell^2}$}, is presented in Fig.~\ref{Figure.skc_bending} as a function of the scale ratio~$L/\ell$ for the two fixed levels of nominal strain. All curves reflect the trends observed in the stress--strain curves. Note that the tangent modulus reduces to approximately~$1/2$ due to pattern transformation in the compressive part.
%
%
\section{Summary and conclusions}
\label{Section.SummaryConclusion}
This paper analyzed size effects in mechanical metamaterials, consisting of hyperelastic materials with periodically arranged circular holes, which exhibit reversible pattern transformations under local or global compressive loading and bending. The study was carried out on a problem with geometrical and material properties based on the experimental \textit{Specimen~1} reported in the work of \cite{Bertoldi2008d}. Infinitely wide specimens with a varying number of holes in the vertical direction ranging between~$ 4 - 128 $ were studied, to obtain a qualitative and quantitative understanding of the size effect. The obtained solutions are largely influenced by the applied boundary conditions, for which two distinct cases were studied. The first case imposed fully constrained top and bottom edges subjected to uniform compression. This resulted in a complete patterned transformation of the deformation field away from the top and bottom boundaries. Here the constrained boundaries induce boundary layers, the thickness of which is influenced by the scale ratio. For specimens with fewer holes along the height direction, the ensemble averaged nominal stresses are largely influenced by the boundary layers. For the cases with a large scale ratio, the influence of the boundary layers was relatively small. The second study involved top and bottom specimen boundaries that are fully unconstrained, for which the lateral edges were rotated to introduce bending deformation. This resulted in partial patterning of the deformation field, where only the bottom half of the specimen transforms (bifurcates). Here, the boundary effect is less prominent, whereas a gradient effect emerges instead.    

For the compression case, the nominal stress~$F/(2\ell)$ is used to quantify the size effect. The thickness of the boundary layer relative to the height of the specimen is studied across the full range of scale ratios. For the bending case, the nominal stress at the top and bottom boundary~$6M/L^2$ is analyzed against the scale ratio. Decreasing the scale ratio in this nonlinear problem has a much more significant influence compared to the linear elastic case studied earlier (see \cite{Ameen2018}). The limitation of classical homogenization in capturing the size effect is also demonstrated. The result shows that for relatively large scale ratios the classical homogenization solution matches well with the ensemble averaged numerical solution, but as the scale ratio reduces, the scale independent classical homogenization solution starts to deviate significantly.

Some key results from this work can be summarized as follows:
\begin{enumerate}

\item For the case of uniaxial compression with fully constrained top and bottom boundaries, the ensemble averaged nominal stress depends strongly on the scale ratio~$L/\ell$. Quantified numerically, the maximum deviation from its homogenized limit exceeds~$ 40\,\% $. Spatially positioning the microstructure is only of limited influence, not exceeding~$5\,\%$.

\item The boundary layer in the case of compression converges with increasing scale ratio to a value of about~$3$ layers of holes close to either boundaries. 

\item For the case of bending with unconstrained top and bottom edges, the ensemble averaged nominal stress remains close to its homogenized limit, the peak difference being less than~$4\,\%$. Spatially positioning the microstructure has a much larger influence, reaching~$50\,\%$ deviation in terms of nominal stress relative to the corresponding homogenized limit. 

\item The flexible boundary layer in the bending case spans again approximately~$3$ layers of holes close to the bottom boundary.

\end{enumerate}

The presented results constitute a rich basis for developing advanced homogenization schemes, to be explored in future work.
%
\appendix
%
%
\section{First-order computational homogenization}
\label{Section:A}
\begin{figure}
	\centering
	\includegraphics[scale=1]{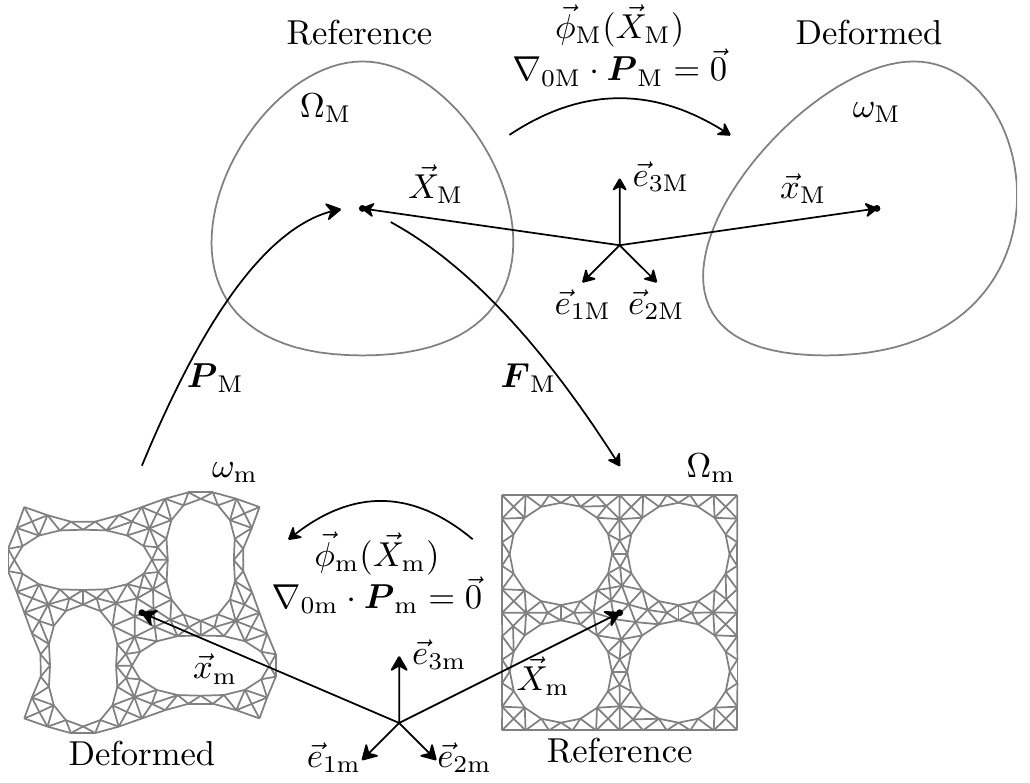}
  	\caption{First-order computational homogenization. The macroscopic kinematic quantities (deformation gradient~$\bs{F}_\mathrm{M}(\vec{X}_\mathrm{M})$) are transferred to the microscale problem defined on an RVE. Resulting microscopic quantities (the first Piola--Kirchhoff stress~$\bs{P}_\mathrm{m}(\vec{X}_\mathrm{m})$) are averaged to provide the macroscopic stress~$\bs{P}_\mathrm{M}(\vec{X}_\mathrm{M})$, a quantity that is transferred back to the macroscale.}
  \label{Figure.A1}
\end{figure}

This section briefly recalls basics of the first-order computational homogenization method, sketched in Fig.~\ref{Figure.A1}, details of which can be found, e.g., in~\citep{Guedes:1990,Kouznetsova:2001}. In general, computational homogenization is a multiscale method that provides a local macroscopic constitutive relation between macroscopic quantities for materials with arbitrary microstructures. That is, it provides the relation
\begin{equation}
\bs{P}_\mathrm{M}(\vec{X}_\mathrm{M}) = \bs{f}(\vec{X}_\mathrm{M},\bs{F}_\mathrm{M}(\vec{X}_\mathrm{M})), \quad \vec{X}_\mathrm{M} \in \Omega_\mathrm{M},
\label{Eq:A1}
\end{equation}
where~$\bs{F}_\mathrm{M}(\vec{X}_\mathrm{M})$ and~$\bs{P}_\mathrm{M}(\vec{X}_\mathrm{M})$ are the macroscopic deformation gradient and the first Piola--Kirchhoff stress tensor respectively, and~$\Omega_\mathrm{M}$ is the macroscopic domain of size~$L$ (recall Section~\ref{SubSect.Geometry}). The constitutive relation~\eqref{Eq:A1} together with the macroscopic balance of linear momentum (in absence of body forces),
\begin{equation}
\nabla_{0\mathrm{M}} \cdot \bs{P}_\mathrm{M}^c(\vec{X}_\mathrm{M}) = \vec{0}, \quad \vec{X}_\mathrm{M} \in \Omega_\mathrm{M},
\label{Eq:A2}
\end{equation}
fully govern the evolution of a macroscopic system. Computational homogenization applies especially in situations in which the macroscopic constitutive relation~\eqref{Eq:A1} is not available in closed form due to, e.g., large deformations, nonconvexities, complex material behavior and history dependence, or complicated microstructural morphology. It relies on the construction and solution of an underlying micro-scale boundary value problem defined on a Representative Volume Element~(RVE), evolution of which is fully governed by the microscopic balance of linear momentum,
\begin{equation}
\nabla_{0\mathrm{m}} \cdot \bs{P}_\mathrm{m}^c(\vec{X}_\mathrm{m}) = \vec{0}, \quad \vec{X}_\mathrm{m} \in \Omega_\mathrm{m}.
\label{Eq:A3}
\end{equation}
In Eq.~\eqref{Eq:A3}, $ \bs{P}_\mathrm{m}(\vec{X}_\mathrm{m}) $ denotes the microscopic first Piloa--Kirchhoff stress tensor, which depends on the microscopic deformation gradient $ \bs{F}_\mathrm{m}(\vec{X}_\mathrm{m}) $, and~$\Omega_\mathrm{m}$ is the microscopic domain with typical microstructural features of size~$\ell$. Dirichlet boundary conditions applied on~$\partial\Omega_\mathrm{m}$ derive from~$\bs{F}_\mathrm{M}$ (cf. e.g.~\citealp[Table~I]{Coenen:2012}) and are usually chosen to be periodic, implying an assumption on local periodicity. The macroscopic quantities (including macroscopic consistent constitutive tangent operator~$\bs{C}^4_\mathrm{M}(\vec{X}_\mathrm{M})$), are identified through averaging of the resulting microscopic quantities, i.e. e.g.
\begin{equation}
\begin{aligned}
\bs{P}_\mathrm{M}(\vec{X}_\mathrm{M}) &= \frac{1}{|\Omega_\mathrm{m}|}\int_{\Omega_\mathrm{m}} \bs{P}_\mathrm{m}(\vec{X}_\mathrm{m}) \, \mathrm{d}\vec{X}_\mathrm{m}, \\
\bs{F}_\mathrm{M}(\vec{X}_\mathrm{M}) &= \frac{1}{|\Omega_\mathrm{m}|}\int_{\Omega_\mathrm{m}} \bs{F}_\mathrm{m}(\vec{X}_\mathrm{m}) \, \mathrm{d}\vec{X}_\mathrm{m},
\end{aligned}
 \quad\quad \vec{X}_\mathrm{M} \in \Omega_\mathrm{M},
\label{Eq:A4}
\end{equation}
which satisfy, along with the boundary conditions applied on~$\partial\Omega_\mathrm{m}$, the macro-homogeneity Hill--Mandel condition
. Constitutive models and morphology at the microscale are assumed to be known, and can be a function of~$\vec{X}_\mathrm{M}$ (meaning effectively that different points can have different RVEs). When both problems~\eqref{Eq:A1} and~\eqref{Eq:A3} are solved simultaneously, a fully nested solution scheme results (sometimes also abbreviated~FE\textsuperscript{2}).

Since the method transfers across the scales only the deformation gradient~$\bs{F}_\mathrm{M}$, it implicitly implies strict separation of scales (i.e.~$L/\ell \rightarrow \infty$) and the standard locality assumption
\begin{equation}
\Delta\vec{x}_\mathrm{m}(\vec{X}_\mathrm{m}) = \bs{F}_\mathrm{M}(\vec{X}_\mathrm{M}) \cdot \Delta\vec{X}_\mathrm{m} + \vec{w}(\vec{X}_\mathrm{m}), \quad \vec{X}_\mathrm{m} \in \Omega_\mathrm{m}, \quad \vec{X}_\mathrm{M} \in \Omega_\mathrm{M},
\label{Eq:A5}
\end{equation}
where~$\vec{w}(\vec{X}_\mathrm{m})$ is the microscopic fluctuation field. Eq.~\eqref{Eq:A5} entails that the method is unable to capture any direct interaction between individual RVEs (i.e. between macroscopic material points~$\vec{X}_\mathrm{M}$) and hence, phenomena such as localization, crack propagation, size effects, boundary layers, and steep strain gradients cannot be accommodated. Extended versions of computational homogenization alleviating some of these limitations involve higher strain gradients or other quantities, see e.g.~\cite{Coenen:2012}.
\begin{figure}
	\centering
	\subfloat[macrostructure and RVEs, $L/h_\mathrm{M} = 4$]{
	\scalebox{0.65}{
	\begin{tikzpicture}[node distance=1em, auto]  
	\linespread{1}
	\tikzset{
    	mynode/.style={rectangle,rounded corners,draw=black, top color=white,inner sep=0.25em,outer sep=0.0em,minimum size=3em,text centered,scale=0.45,transform shape},
	    myarrow/.style={->, >=latex', shorten >=1pt},
	}  
	
	\node[inner sep=0em,outer sep=0em] (Macro) {
		\includegraphics[scale=1]{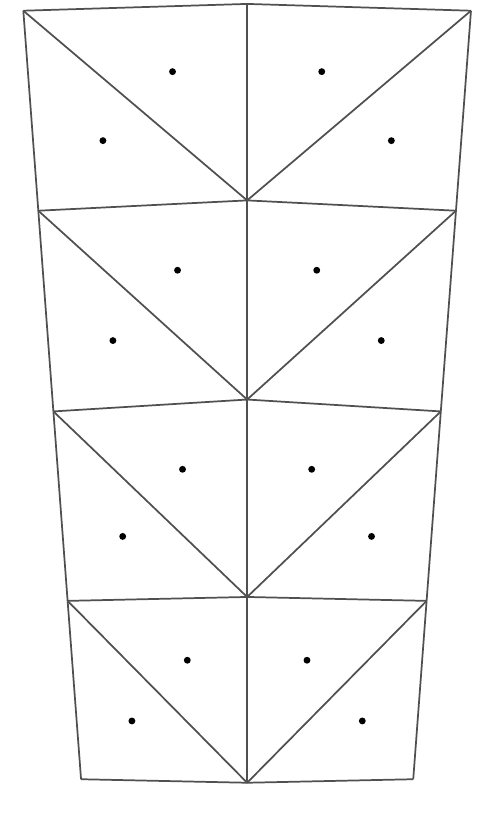}
	};
			
	\node[inner sep=0em,outer sep=0em,above right=2.0em of Macro, shift={(0,-3.4)}] (RVE1) {
		\includegraphics[scale=0.5]{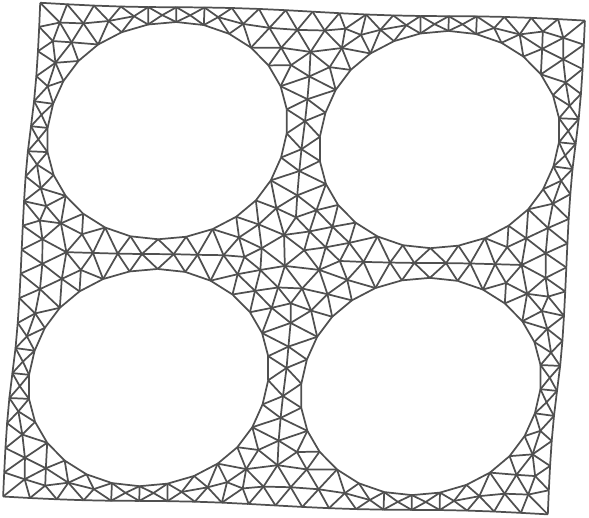}
	};
	\node[inner sep=0em,outer sep=0em,left=2.0em of Macro, shift={(0,1.6)}] (RVE2) {
		\includegraphics[scale=0.5]{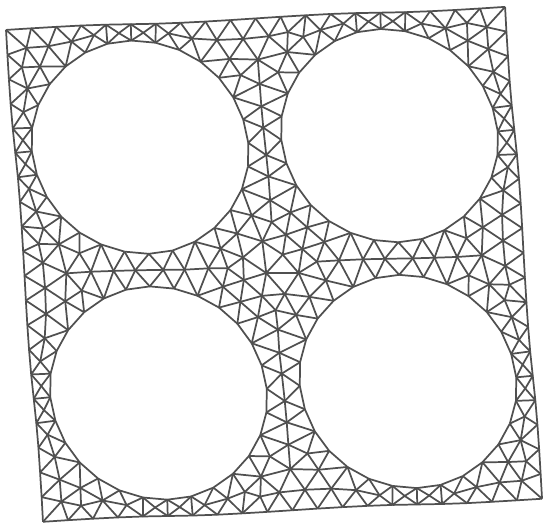}
	}; 
	\node[inner sep=0em,outer sep=0em,right=1.0em of Macro, shift={(0,-1.6)}] (RVE3) {
		\includegraphics[scale=0.5]{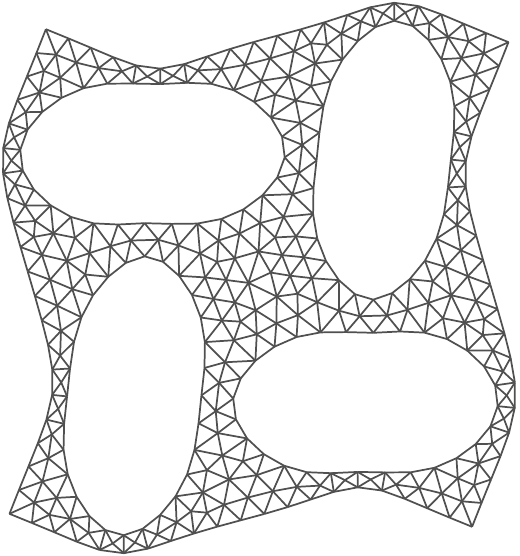}
	};
	\node[inner sep=0em,outer sep=0em,left=2.0em of Macro, shift={(0,-2.5)}] (RVE4) {
		\includegraphics[scale=0.5]{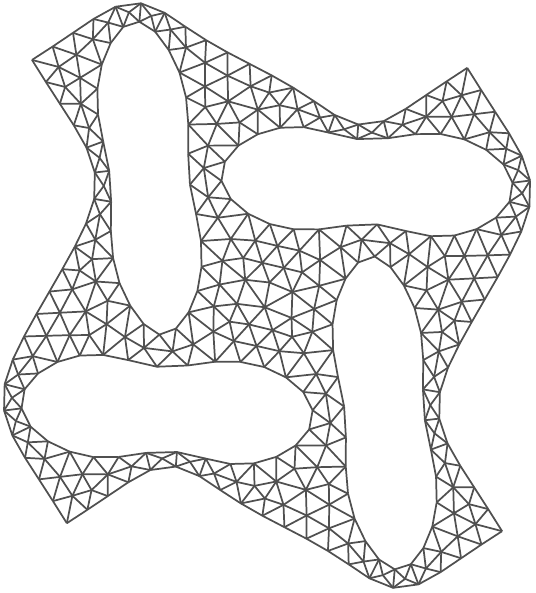}
	};
	
 	\draw[black, dashed] (1.5,2.83) -- (RVE1.north west);
 	\draw[black, dashed] (1.5,2.83) -- (RVE1.south west);
	 	
 	\draw[black, dashed] (-1.35,0.8) -- (RVE2.north east);
 	\draw[black, dashed] (-1.35,0.8) -- (RVE2.south east);
	 	
 	\draw[black, dashed] (1.28,-1.2) -- (RVE3.north west);
 	\draw[black, dashed] (1.28,-1.2) -- (RVE3.south west);

 	\draw[black, dashed] (-1.15,-3.08) -- (RVE4.north east);
 	\draw[black, dashed] (-1.15,-3.08) -- (RVE4.south east);	 		 		 	
		 	 	 	
		
	\end{tikzpicture}
	}\label{Figure.A2a}}\hspace{1em}
	\subfloat[nominal stress versus strain]{
	\includegraphics[scale=1]{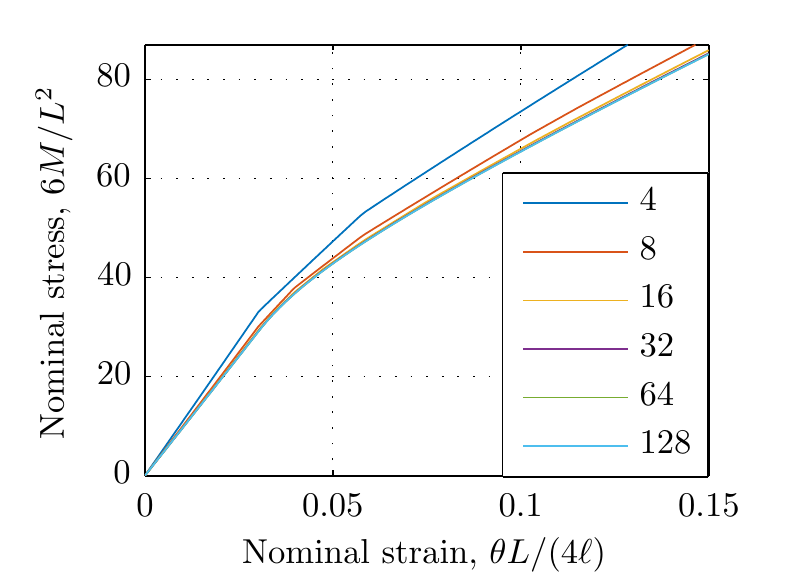}\label{Figure.A2b}
	}
  	\caption{First-order computational homogenization in the case of bending. (a)~Deformed macrostructure and four RVEs, and~(b) macroscopic mesh convergence study, i.e. dependence of nominal stress--strain curves on various sizes of the macroscopic mesh elements~$L/h_\mathrm{M} = 4, \dots, 128$.}
  \label{Figure.A2}
\end{figure}
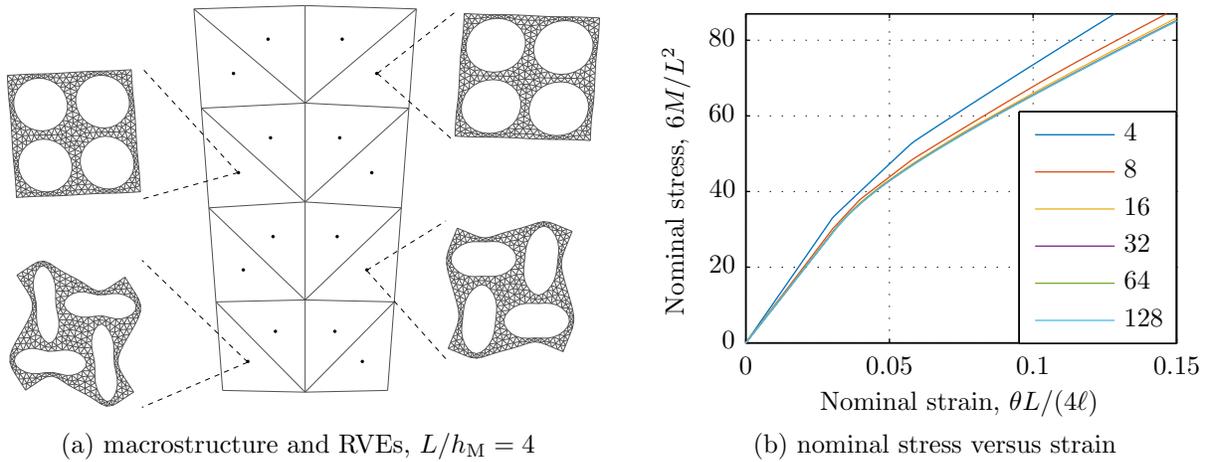

In the context of this paper, the microscopic stress is obtained through differentiation of the strain energy density~$W$ (recall Eq.~\eqref{Eq:1}) with respect to the microscopic deformation gradient~$\bs{F}_\mathrm{m}(\vec{X}_\mathrm{m})$, i.e.
\begin{equation}
P^\mathrm{m}_{ij}(\vec{X}_\mathrm{m}) = \frac{\partial W(\bs{F}_\mathrm{m}(\vec{X}_\mathrm{m}))}{\partial F^\mathrm{m}_{ij}}, \quad \vec{X}_\mathrm{m} \in \Omega_\mathrm{m}.
\label{Eq:A6}
\end{equation}
In Eq.~\eqref{Eq:A6} the subscript~``$\mathrm{m}$" is raised to a superscript wherever the indicial notation has been used. The RVE morphology is considered constant in~$\vec{X}_\mathrm{M}$, chosen to be~$2 \times 2$ unit cells large based on prior insights from full scale simulations, although in general identification of the correct size is a rather delicate problem on its own, cf. e.g.~\cite{Saiki:2002}. Boundary conditions applied on~$\partial\Omega_\mathrm{m}$ are chosen to be periodic. As a consequence of the separation of scales, the strain field obtained from the first-order computational homogenization is constant in the case of compression (recall Section~\ref{UniaxCompression}), ignoring any boundary layers. This means that all macroscopic points~$\vec{X}_\mathrm{M}$ (and hence RVEs) experience the same state of deformation and only one RVE suffices to carry out the simulation. In the case of bending the situation changes due to the presence of a strain gradient, cf. Fig.~\ref{Figure.A2} and Section~\ref{Bending}, meaning that individual RVEs experience different states of deformation. In order to avoid any bias due to discretization, a mesh sensitivity study has been performed to identify the macroscopic element size~$h_\mathrm{M}$ that yields accurate results ($128$ elements per specimen height~$L$), cf. also Fig.~\ref{Figure.A2b}.
%
%
\section*{Acknowledgements}
The research leading to these results has received funding from the European Research Council under the European Union's Seventh Framework Programme (FP7/2007-2013)/ERC grant agreement \textnumero~[339392].
%
%
%


\begin{thebibliography}{27}
	\expandafter\ifx\csname natexlab\endcsname\relax\def\natexlab#1{#1}\fi
	\providecommand{\url}[1]{\texttt{#1}}
	\providecommand{\href}[2]{#2}
	\providecommand{\path}[1]{#1}
	\providecommand{\DOIprefix}{doi: }
	\providecommand{\ArXivprefix}{arXiv: }
	\providecommand{\URLprefix}{URL: }
	\providecommand{\Pubmedprefix}{pmid: }
	\providecommand{\doi}[1]{\href{http://dx.doi.org/#1}{\path{#1}}}
	\providecommand{\Pubmed}[1]{\href{pmid:#1}{\path{#1}}}
	\providecommand{\bibinfo}[2]{#2}
	\ifx\xfnm\relax \def\xfnm[#1]{\unskip,\space#1}\fi
	\bibitem[{Ameen et~al.(2018)Ameen, Peerlings and Geers}]{Ameen2018}
	\bibinfo{author}{Ameen, M.M.}, \bibinfo{author}{Peerlings, R.H.J.},
	\bibinfo{author}{Geers, M.G.D.}, \bibinfo{year}{2018}.
	\newblock \bibinfo{title}{A quantitative assessment of the scale separation
		limits of classical and higher-order asymptotic homogenization}.
	\newblock \bibinfo{journal}{European Journal of Mechanics-A/Solids}
	\bibinfo{volume}{71}, \bibinfo{pages}{89--100}.
	\newblock \URLprefix
	\url{https://www.sciencedirect.com/science/article/pii/S0997753817303303},
	\DOIprefix\doi{10.1016/j.euromechsol.2018.02.011}.
	\bibitem[{Arzt(1998)}]{arzt1998}
	\bibinfo{author}{Arzt, E.}, \bibinfo{year}{1998}.
	\newblock \bibinfo{title}{Size effects in materials due to microstructural and
		dimensional constraints: a comparative review}.
	\newblock \bibinfo{journal}{Acta Materialia} \bibinfo{volume}{46},
	\bibinfo{pages}{5611--5626}.
	\newblock \URLprefix
	\url{http://www.sciencedirect.com/science/article/pii/S1359645498002316},
	\DOIprefix\doi{http://dx.doi.org/10.1016/S1359-6454(98)00231-6}.
	\bibitem[{Bertoldi et~al.(2008)Bertoldi, Boyce, Deschanel, Prange and
		Mullin}]{Bertoldi2008d}
	\bibinfo{author}{Bertoldi, K.}, \bibinfo{author}{Boyce, M.C.},
	\bibinfo{author}{Deschanel, S.}, \bibinfo{author}{Prange, S.M.},
	\bibinfo{author}{Mullin, T.}, \bibinfo{year}{2008}.
	\newblock \bibinfo{title}{Mechanics of deformation-triggered pattern
		transformations and superelastic behavior in periodic elastomeric
		structures}.
	\newblock \bibinfo{journal}{Journal of the Mechanics and Physics of Solids}
	\bibinfo{volume}{56}, \bibinfo{pages}{2642--2668}.
	\newblock \URLprefix
	\url{http://www.sciencedirect.com/science/article/pii/S0022509608000434},
	\DOIprefix\doi{http://dx.doi.org/10.1016/j.jmps.2008.03.006}.
	\bibitem[{Bertoldi et~al.(2010)Bertoldi, Reis, Willshaw and
		Mullin}]{Bertoldi2010a}
	\bibinfo{author}{Bertoldi, K.}, \bibinfo{author}{Reis, P.M.},
	\bibinfo{author}{Willshaw, S.}, \bibinfo{author}{Mullin, T.},
	\bibinfo{year}{2010}.
	\newblock \bibinfo{title}{Negative poisson's ratio behavior induced by an
		elastic instability}.
	\newblock \bibinfo{journal}{Advanced Materials} \bibinfo{volume}{22},
	\bibinfo{pages}{361--366}.
	\newblock \URLprefix \url{http://dx.doi.org/10.1002/adma.200901956},
	\DOIprefix\doi{10.1002/adma.200901956}.
	\bibitem[{de~Borst et~al.(2012)de~Borst, Crisfield, Remmers and
		Verhoosel}]{crisfield2012}
	\bibinfo{author}{de~Borst, R.}, \bibinfo{author}{Crisfield, M.A.},
	\bibinfo{author}{Remmers, J.J.C.}, \bibinfo{author}{Verhoosel, C.V.},
	\bibinfo{year}{2012}.
	\newblock \bibinfo{title}{Nonlinear finite element analysis of solids and
		structures}.
	\newblock \bibinfo{publisher}{John Wiley \& Sons}.
	\newblock \DOIprefix\doi{10.1002/9781118375938}.
	\bibitem[{Brezny and Green(1990)}]{brezny1990}
	\bibinfo{author}{Brezny, R.}, \bibinfo{author}{Green, D.J.},
	\bibinfo{year}{1990}.
	\newblock \bibinfo{title}{The effect of cell size on the mechanical behavior of
		cellular materials}.
	\newblock \bibinfo{journal}{Acta Metallurgica et Materialia}
	\bibinfo{volume}{38}, \bibinfo{pages}{2517--2526}.
	\newblock \URLprefix
	\url{http://www.sciencedirect.com/science/article/pii/095671519090263G},
	\DOIprefix\doi{http://dx.doi.org/10.1016/0956-7151(90)90263-G}.
	\bibitem[{Chen and Fleck(2002)}]{chen2002size}
	\bibinfo{author}{Chen, C.}, \bibinfo{author}{Fleck, N.A.},
	\bibinfo{year}{2002}.
	\newblock \bibinfo{title}{Size effects in the constrained deformation of
		metallic foams}.
	\newblock \bibinfo{journal}{Journal of the Mechanics and Physics of Solids}
	\bibinfo{volume}{50}, \bibinfo{pages}{955--977}.
	\newblock \URLprefix
	\url{http://www.sciencedirect.com/science/article/pii/S0022509601001284},
	\DOIprefix\doi{http://dx.doi.org/10.1016/S0022-5096(01)00128-4}.
	\bibitem[{Coenen et~al.(2012)Coenen, Kouznetsova and Geers}]{Coenen:2012}
	\bibinfo{author}{Coenen, E.}, \bibinfo{author}{Kouznetsova, V.},
	\bibinfo{author}{Geers, M.}, \bibinfo{year}{2012}.
	\newblock \bibinfo{title}{Novel boundary conditions for strain localization
		analyses in microstructural volume elements}.
	\newblock \bibinfo{journal}{International Journal for Numerical Methods in
		Engineering} \bibinfo{volume}{90}, \bibinfo{pages}{1--21}.
	\newblock \URLprefix \url{http://dx.doi.org/10.1002/nme.3298},
	\DOIprefix\doi{10.1002/nme.3298}.
	\bibitem[{Coulais(2016)}]{Coulais2016a}
	\bibinfo{author}{Coulais, C.}, \bibinfo{year}{2016}.
	\newblock \bibinfo{title}{Periodic cellular materials with nonlinear elastic
		homogenized stress-strain response at small strains}.
	\newblock \bibinfo{journal}{International Journal of Solids and Structures}
	\bibinfo{volume}{97}, \bibinfo{pages}{226--238}.
	\newblock \URLprefix
	\url{http://www.sciencedirect.com/science/article/pii/S002076831630186X},
	\DOIprefix\doi{http://dx.doi.org/10.1016/j.ijsolstr.2016.07.025}.
	\bibitem[{Florijn et~al.(2014)Florijn, Coulais and van Hecke}]{florijn2014}
	\bibinfo{author}{Florijn, B.}, \bibinfo{author}{Coulais, C.},
	\bibinfo{author}{van Hecke, M.}, \bibinfo{year}{2014}.
	\newblock \bibinfo{title}{Programmable mechanical metamaterials}.
	\newblock \bibinfo{journal}{Phys. Rev. Lett.} \bibinfo{volume}{113},
	\bibinfo{pages}{175503}.
	\newblock \URLprefix
	\url{https://link.aps.org/doi/10.1103/PhysRevLett.113.175503},
	\DOIprefix\doi{10.1103/PhysRevLett.113.175503}.
	\bibitem[{Frantziskonis et~al.(1997)Frantziskonis, Renaudin and
		Breysse}]{frantziskonis1997}
	\bibinfo{author}{Frantziskonis, G.}, \bibinfo{author}{Renaudin, P.},
	\bibinfo{author}{Breysse, D.}, \bibinfo{year}{1997}.
	\newblock \bibinfo{title}{Heterogeneous solids - {P}art {I}: analytical and
		numerical 1-{D} results on boundary effects}.
	\newblock \bibinfo{journal}{European Journal of Mechanics, A/Solids}
	\bibinfo{volume}{16}, \bibinfo{pages}{409--423}.
	\bibitem[{Geuzaine and Remacle(2009)}]{NME2579}
	\bibinfo{author}{Geuzaine, C.}, \bibinfo{author}{Remacle, J.F.},
	\bibinfo{year}{2009}.
	\newblock \bibinfo{title}{Gmsh: A 3-d finite element mesh generator with
		built-in pre-and post-processing facilities}.
	\newblock \bibinfo{journal}{International journal for numerical methods in
		engineering} \bibinfo{volume}{79}, \bibinfo{pages}{1309--1331}.
	\newblock \URLprefix \url{http://dx.doi.org/10.1002/nme.2579},
	\DOIprefix\doi{10.1002/nme.2579}.
	\bibitem[{Geymonat et~al.(1993)Geymonat, M{\"u}ller and
		Triantafyllidis}]{geymonat1993}
	\bibinfo{author}{Geymonat, G.}, \bibinfo{author}{M{\"u}ller, S.},
	\bibinfo{author}{Triantafyllidis, N.}, \bibinfo{year}{1993}.
	\newblock \bibinfo{title}{Homogenization of nonlinearly elastic materials,
		microscopic bifurcation and macroscopic loss of rank-one convexity}.
	\newblock \bibinfo{journal}{Archive for Rational Mechanics and Analysis}
	\bibinfo{volume}{122}, \bibinfo{pages}{231--290}.
	\newblock \DOIprefix\doi{http://dx.doi.org/10.1007/BF00380256}.
	\bibitem[{Guedes and Kikuchi(1990)}]{Guedes:1990}
	\bibinfo{author}{Guedes, J.M.}, \bibinfo{author}{Kikuchi, N.},
	\bibinfo{year}{1990}.
	\newblock \bibinfo{title}{Preprocessing and postprocessing for materials based
		on the homogenization method with adaptive finite element methods}.
	\newblock \bibinfo{journal}{Computer Methods in Applied Mechanics and
		Engineering} \bibinfo{volume}{83}, \bibinfo{pages}{143 -- 198}.
	\newblock \URLprefix
	\url{http://www.sciencedirect.com/science/article/pii/004578259090148F},
	\DOIprefix\doi{http://dx.doi.org/10.1016/0045-7825(90)90148-F}.
	\bibitem[{Hu et~al.(2013)Hu, He, Lei and Liu}]{Hu2013}
	\bibinfo{author}{Hu, J.}, \bibinfo{author}{He, Y.}, \bibinfo{author}{Lei, J.},
	\bibinfo{author}{Liu, Z.}, \bibinfo{year}{2013}.
	\newblock \bibinfo{title}{Novel mechanical behavior of periodic structure with
		the pattern transformation}.
	\newblock \bibinfo{journal}{Theoretical and Applied Mechanics Letters}
	\bibinfo{volume}{3}.
	\newblock \URLprefix
	\url{http://www.sciencedirect.com/science/article/pii/S2095034915302634},
	\DOIprefix\doi{http://dx.doi.org/10.1063/2.1305407}.
	\bibitem[{Kouznetsova et~al.(2001)Kouznetsova, Brekelmans and
		Baaijens}]{Kouznetsova:2001}
	\bibinfo{author}{Kouznetsova, V.}, \bibinfo{author}{Brekelmans, W.A.M.},
	\bibinfo{author}{Baaijens, F.P.T.}, \bibinfo{year}{2001}.
	\newblock \bibinfo{title}{An approach to micro-macro modeling of heterogeneous
		materials}.
	\newblock \bibinfo{journal}{Computational Mechanics} \bibinfo{volume}{27},
	\bibinfo{pages}{37--48}.
	\newblock \URLprefix \url{https://doi.org/10.1007/s004660000212},
	\DOIprefix\doi{10.1007/s004660000212}.
	\bibitem[{Krishnan and Johnson(2009)}]{Krishnan2009}
	\bibinfo{author}{Krishnan, D.}, \bibinfo{author}{Johnson, H.},
	\bibinfo{year}{2009}.
	\newblock \bibinfo{title}{Optical properties of two-dimensional polymer
		photonic crystals after deformation-induced pattern transformations}.
	\newblock \bibinfo{journal}{Journal of the Mechanics and Physics of Solids}
	\bibinfo{volume}{57}, \bibinfo{pages}{1500--1513}.
	\newblock \URLprefix
	\url{http://www.sciencedirect.com/science/article/pii/S0022509609000878},
	\DOIprefix\doi{http://dx.doi.org/10.1016/j.jmps.2009.05.012}.
	\bibitem[{Liu et~al.(2016)Liu, Gu, Shan, Kang, Weaver and Bertoldi}]{Liu2016a}
	\bibinfo{author}{Liu, J.}, \bibinfo{author}{Gu, T.}, \bibinfo{author}{Shan,
		S.}, \bibinfo{author}{Kang, S.H.}, \bibinfo{author}{Weaver, J.C.},
	\bibinfo{author}{Bertoldi, K.}, \bibinfo{year}{2016}.
	\newblock \bibinfo{title}{Harnessing buckling to design architected materials
		that exhibit effective negative swelling}.
	\newblock \bibinfo{journal}{Advanced Materials} \bibinfo{volume}{28},
	\bibinfo{pages}{6619--6624}.
	\newblock \URLprefix \url{http://dx.doi.org/10.1002/adma.201600812},
	\DOIprefix\doi{10.1002/adma.201600812}.
	\bibitem[{Mullin et~al.(2007)Mullin, Deschanel, Bertoldi and
		Boyce}]{Mullin2007a}
	\bibinfo{author}{Mullin, T.}, \bibinfo{author}{Deschanel, S.},
	\bibinfo{author}{Bertoldi, K.}, \bibinfo{author}{Boyce, M.C.},
	\bibinfo{year}{2007}.
	\newblock \bibinfo{title}{Pattern transformation triggered by deformation}.
	\newblock \bibinfo{journal}{Physical Review Letters} \bibinfo{volume}{99},
	\bibinfo{pages}{084301}.
	\newblock \DOIprefix\doi{10.1103/PhysRevLett.99.084301}.
	\bibitem[{Onck et~al.(2001)Onck, Andrews and Gibson}]{onck2001size}
	\bibinfo{author}{Onck, P.R.}, \bibinfo{author}{Andrews, E.W.},
	\bibinfo{author}{Gibson, L.J.}, \bibinfo{year}{2001}.
	\newblock \bibinfo{title}{Size effects in ductile cellular solids. part {I}:
		modeling}.
	\newblock \bibinfo{journal}{International Journal of Mechanical Sciences}
	\bibinfo{volume}{43}, \bibinfo{pages}{681--699}.
	\newblock \URLprefix
	\url{http://www.sciencedirect.com/science/article/pii/S0020740300000424},
	\DOIprefix\doi{http://dx.doi.org/10.1016/S0020-7403(00)00042-4}.
	\bibitem[{Saiki et~al.(2002)Saiki, Terada, Ikeda and Hori}]{Saiki:2002}
	\bibinfo{author}{Saiki, I.}, \bibinfo{author}{Terada, K.},
	\bibinfo{author}{Ikeda, K.}, \bibinfo{author}{Hori, M.},
	\bibinfo{year}{2002}.
	\newblock \bibinfo{title}{Appropriate number of unit cells in a representative
		volume element for micro-structural bifurcation encountered in a multi-scale
		modeling}.
	\newblock \bibinfo{journal}{Computer Methods in Applied Mechanics and
		Engineering} \bibinfo{volume}{191}, \bibinfo{pages}{2561 -- 2585}.
	\newblock \URLprefix
	\url{http://www.sciencedirect.com/science/article/pii/S0045782501004133},
	\DOIprefix\doi{http://dx.doi.org/10.1016/S0045-7825(01)00413-3}.
	\bibitem[{Shan et~al.(2014)Shan, Kang, Wang, Qu, Shian, Chen and
		Bertoldi}]{Shan2014a}
	\bibinfo{author}{Shan, S.}, \bibinfo{author}{Kang, S.H.},
	\bibinfo{author}{Wang, P.}, \bibinfo{author}{Qu, C.}, \bibinfo{author}{Shian,
		S.}, \bibinfo{author}{Chen, E.R.}, \bibinfo{author}{Bertoldi, K.},
	\bibinfo{year}{2014}.
	\newblock \bibinfo{title}{Harnessing multiple folding mechanisms in soft
		periodic structures for tunable control of elastic waves}.
	\newblock \bibinfo{journal}{Advanced Functional Materials}
	\bibinfo{volume}{24}, \bibinfo{pages}{4935--4942}.
	\newblock \URLprefix \url{http://dx.doi.org/10.1002/adfm.201400665},
	\DOIprefix\doi{10.1002/adfm.201400665}.
	\bibitem[{Singamaneni et~al.(2008)Singamaneni, Bertoldi, Chang, Jang, Thomas,
		Boyce and Tsukruk}]{Singamaneni2009a}
	\bibinfo{author}{Singamaneni, S.}, \bibinfo{author}{Bertoldi, K.},
	\bibinfo{author}{Chang, S.}, \bibinfo{author}{Jang, J.H.},
	\bibinfo{author}{Thomas, E.L.}, \bibinfo{author}{Boyce, M.C.},
	\bibinfo{author}{Tsukruk, V.V.}, \bibinfo{year}{2008}.
	\newblock \bibinfo{title}{Instabilities and pattern transformation in periodic,
		porous elastoplastic solid coatings}.
	\newblock \bibinfo{journal}{ACS Applied Materials \& Interfaces}
	\bibinfo{volume}{1}, \bibinfo{pages}{42--47}.
	\newblock \DOIprefix\doi{10.1021/am800078f}.
	\bibitem[{Singamaneni and Tsukruk(2010)}]{Singamaneni2010a}
	\bibinfo{author}{Singamaneni, S.}, \bibinfo{author}{Tsukruk, V.V.},
	\bibinfo{year}{2010}.
	\newblock \bibinfo{title}{Buckling instabilities in periodic composite
		polymeric materials}.
	\newblock \bibinfo{journal}{Soft Matter} \bibinfo{volume}{6},
	\bibinfo{pages}{5681--5692}.
	\newblock \URLprefix \url{http://dx.doi.org/10.1039/C0SM00374C},
	\DOIprefix\doi{10.1039/C0SM00374C}.
	\bibitem[{Smyshlyaev and Cherednichenko(2000)}]{Smyshlyaev}
	\bibinfo{author}{Smyshlyaev, V.P.}, \bibinfo{author}{Cherednichenko, K.D.},
	\bibinfo{year}{2000}.
	\newblock \bibinfo{title}{On rigorous derivation of strain gradient effects in
		the overall behaviour of periodic heterogeneous media}.
	\newblock \bibinfo{journal}{Journal of the Mechanics and Physics of Solids}
	\bibinfo{volume}{48}, \bibinfo{pages}{1325--1357}.
	\newblock \URLprefix
	\url{http://www.sciencedirect.com/science/article/pii/S0022509699000903},
	\DOIprefix\doi{http://dx.doi.org/10.1016/S0022-5096(99)00090-3}.
	\bibitem[{Wriggers(2008)}]{Wriggers2008b}
	\bibinfo{author}{Wriggers, P.}, \bibinfo{year}{2008}.
	\newblock \bibinfo{title}{Nonlinear finite element methods}.
	\newblock \bibinfo{publisher}{Springer Science \& Business Media}.
	\bibitem[{Yang et~al.(2015)Yang, Mosadegh, Ainla, Lee, Khashai, Suo, Bertoldi
		and Whitesides}]{Yang2015}
	\bibinfo{author}{Yang, D.}, \bibinfo{author}{Mosadegh, B.},
	\bibinfo{author}{Ainla, A.}, \bibinfo{author}{Lee, B.},
	\bibinfo{author}{Khashai, F.}, \bibinfo{author}{Suo, Z.},
	\bibinfo{author}{Bertoldi, K.}, \bibinfo{author}{Whitesides, G.M.},
	\bibinfo{year}{2015}.
	\newblock \bibinfo{title}{Buckling of elastomeric beams enables actuation of
		soft machines}.
	\newblock \bibinfo{journal}{Advanced Materials} \bibinfo{volume}{27},
	\bibinfo{pages}{6323--6327}.
	\newblock \URLprefix \url{http://dx.doi.org/10.1002/adma.201503188},
	\DOIprefix\doi{10.1002/adma.201503188}.
	
\end{thebibliography}

\end{document}